\documentclass[10pt,twocolumn]{article}
\usepackage[dvips]{graphicx}
\DeclareGraphicsExtensions{ps,eps}
\usepackage{latexsym}
\def\beq{\begin{equation}}
\def\eeq{\end{equation}}

\begin{document}
\twocolumn[
\begin{center}
{\bf \Large Cross-spectral analysis of physiological\\
                 tremor and muscle activity.\\
            II. Application to synchronized EMG\\} \vskip 0.4cm
     {      J. Timmer${}^{1}$,
                M. Lauk${}^{1,2}$,
                W. Pfleger${}^{1}$ and
                G. Deuschl${}^{3}$} \\ \vskip 0.3cm
        {\normalsize ${}^{1}$Zentrum f\"ur Datenanalyse und Modellbildung,
                Eckerstr. 1, 79104 Freiburg, Germany \\              
               ${}^{2}$Neurologische Universit\"atsklinik Freiburg,
               Breisacher Str. 64, 79110 Freiburg, Germany \\
                ${}^{3}$Neurologische Universit\"atsklinik Kiel,
                Niemannsweg 147, 24105 Kiel, Germany} \\ \vskip 0.3cm
        {\small \em (to appear in Biological Cybernetics)}\\ \vskip 0.1cm
\end{center}
\noindent
{\bf We investigate the relationship between synchronized muscle activity
and tremor time series in (enhanced) physiological tremor by 
cross-spectral analysis. 
Special attention is directed to the phase spectrum and its
possibilities to clarify the contribution of reflex mechanisms to 
physiological tremor.
The phase spectra are investigated under the assumptions that the EMG
synchronization was caused by a reflex, respectively a central
oscillator. In comparison of these results to phase spectra of
measured data we found a significant contribution of reflexes.
But reflexes only modify existing peaks in the power spectrum.
The main agents of physiological tremor are an efferent pace
and the resonant behavior of the biomechanical system.
} \vskip .4cm]

\section{Introduction}
The contribution of reflex loops to (enhanced) physiological tremor is
 still a matter
of debate (Elble and Koller 1990). Most experiments investigating
this topic used some kind of manipulation of the system. 
For example, Lippold (1971)
cooled the limb in order to reduce the velocity of nerve
conduction, Rack et al.~(1978) examined the reflex response to
enforced sinusoidal movement of the limb and Young and Hagbarth (1980)
investigated several manoeuvres that affect the stretch reflex to
enhance physiological tremor (ETP). We intend to investigate
the role of reflex mechanisms by cross-spectral analysis of surface 
electromyogram (EMG) and 
 accelerometer data (ACC) without perturbing the system
externally. Former applications of 
this technique to physiological tremor are given in (Fox and Randall 1970; 
Pashda and Stein 1973; Elble and Randall 1976; Iaizzo and Pozos 1992). 
These authors interpreted the phase spectrum at single, fixed frequencies
in terms of delays between the EMG and the AC.C
We examine the whole phase spectrum and investigate whether
the phase spectrum can help to clarify the role of reflexes in EPT.

Typical EPT usually shows two significant peaks in the
ACC power spectrum. These peaks exhibit a different behavior if the hand is
loaded. While often indistinguishable located around 10 to 12 Hz if
the hand is not loaded, one peak moves to lower frequencies under
increasing loads. This peak is called mechanical peak since it depends
on the mechanical properties of the musculosceletal system. An EMG peak at 
the same frequency shows up occasionally. The frequency of the second peak 
does not vary under moderate loading and is always accompanied by a peak 
in the EMG power spectrum
(H\"omberg et al.~1987; Elble and Koller 1990). This peak is called
neurogenic. It is still
controversial whether these EMG synchronizations result from reflex loops,
by chance synchronization
or from a central oscillator (see Allum et al.~1978; Stiles 1980; 
Christakos 1982; Allum 1984; 
Elble and Koller 1990 for a review). The two latter hypotheses can not
be distinguished by cross-spectral analysis.

For both hypotheses of reflex mechanisms and non-reflex mechanisms, 
we calculate the power spectrum, the cross-correlation function 
and the phase spectrum.
In the case of the non-reflex hypothesis, it is possible to derive these 
quantities analytically. In the case of reflexes, some kind of 
nonlinearity has to be introduced into the feedback system 
which hinders an analytical treatment.
Based on a stochastic feedback system for the effects of reflexes we 
perform simulation studies to understand the behavior of the 
cross-correlation function and the spectra.
This model is a generalization of the deterministic model introduced by Stein
and Oguzt\"oreli (Stein and Oguzt\"oreli 1975, 1978; Oguzt\"oreli and
Stein 1975, 1976, 1979; Bawa et al.~1976a, 1976b).

It should be mentioned that the well developed theory of Volterra - 
Wiener kernel estimators, which has been successfully applied to 
model physiological feedback systems (Marmarelis 1989) can not be applied
to the problem under investigation since it requires the input data of 
the system, which are not available here. 

We do not intend to derive a quantitative model of 
the reflex mechanism. We will use the phase spectra as an indicator
for the presence of reflex contributions by 
comparing the predicted phase spectra under both hypotheses
 with those estimated from measured data.
For both cases the mechanical properties of the hand are modeled by
an autoregressive process of order to two (AR[2]) covered by
observational noise (Stiles and Randall 1967; 
Randall 1973; Rietz and Stiles 1974; Elble and Koller 1990; 
Gantert et al.~1992). The synchronized (possible) non-reflex part of 
the rectified EMG 
is also described by an AR[2] process. We claim no physiological 
significance for this EMG model; it is used solely as a description of 
the data and an approximation to the true statistical behavior 
of this part of the EMG. Since we only investigate second order statistical 
properties of the signals this description is adequate if the power 
spectra of the model and the data are similar.
This model is justified by the over-all consistent description of the 
phenomena observed in the data.

In a companion paper (Timmer et al.~1998), we have described the
recording conditions, the mathematical methods
and their application to EMG and ACC data for cases where 
the EMG does not show significant
synchronization.

This paper is organized as follows. 
In the next section we discuss the characteristic features of the 
phase spectrum between EMG and ACC assuming that the 
synchronized EMG does not result from a reflex mechanism.
In Section \ref {reflexes} we give the corresponding results 
assuming a significant contribution of reflex loops.
In Section \ref {results} we give the results for 35 data sets
recorded from 19 subjects.

\section {Non reflex mechanisms} \label{central_osci}
In this section, we show the expected behavior of the different kinds
of spectra assuming that the synchronized EMG activity is not caused by any
reflex loop. Such a synchronization can result from motor units firing
asynchronously at similar rates
(Christakos 1982) or some central oscillator. Cross-spectral
analysis will not be able to distinguish between peripheral and
central non-reflex mechanisms.
We first show results of simulations and then applications to measured data.
\subsection {Simulations}

If the synchronization of the EMG is not caused by a reflex, the 
position of the hand, $ x(t) $ (the ACC is obtained by
differentiating $x(t)$ twice), is described 
by an AR process driven by colored noise. 
Denoting uncorrelated EMG activity by $ \nu(t)$, the synchronized EMG by 
$ y(t)$, and a possible time delay of the effect
of the EMG on the ACC by $\Delta t$, the model reads:
\begin{eqnarray} 
y(t) & = & b_1 y(t-1) + b_2 y(t-2) + \epsilon(t) \label{nr1} \\
x(t) & = & a_1 x(t-1) + a_2 x(t-2) \\\nonumber
        &&      \qquad\qquad\qquad + \nu(t-\Delta t)  + y(t- \Delta t) 
        \quad .\label{nr2}
\end{eqnarray}

\begin{figure}[!ht]
\begin{center}
\includegraphics[scale=0.38]{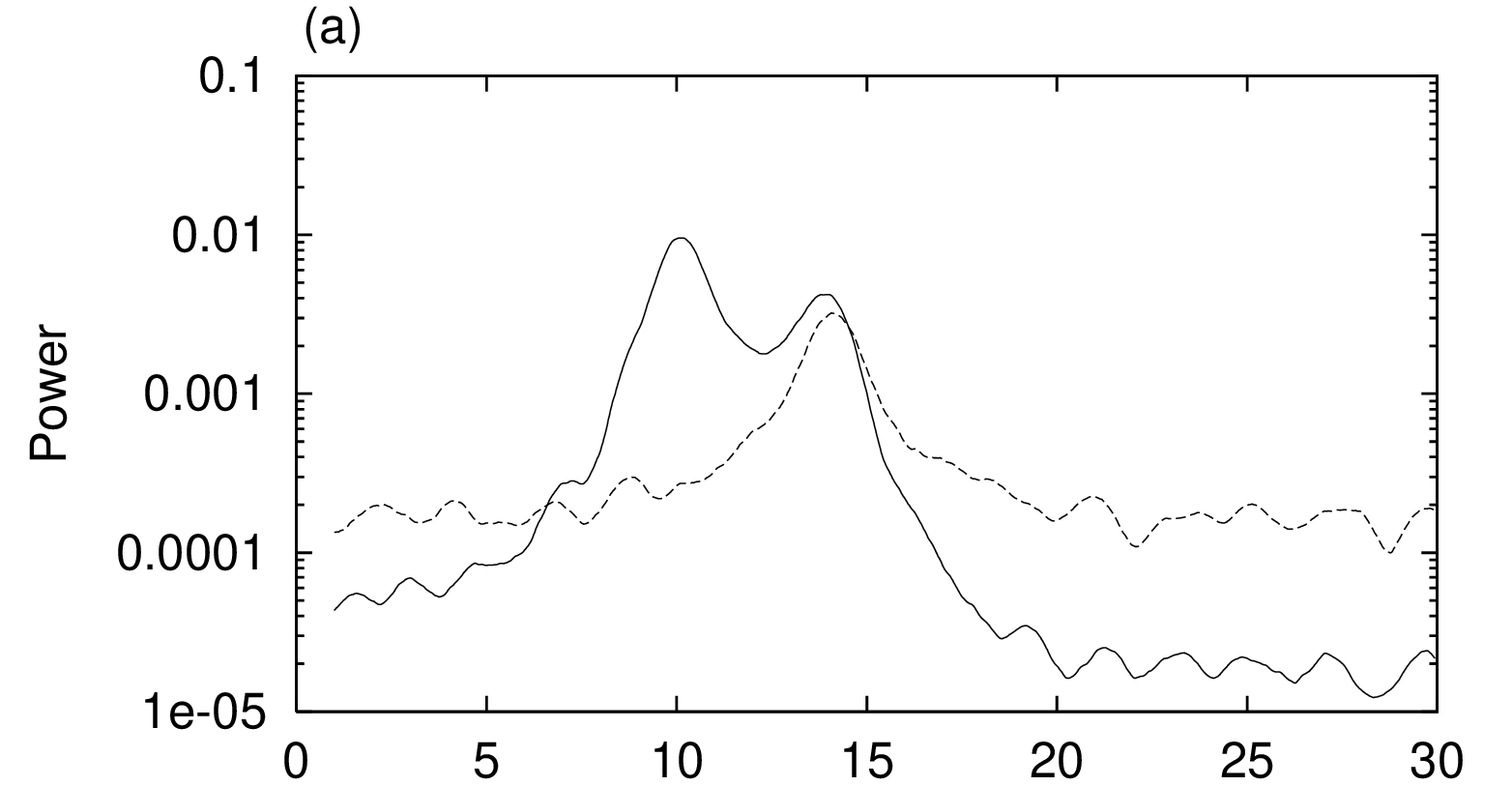}
\includegraphics[scale=0.38]{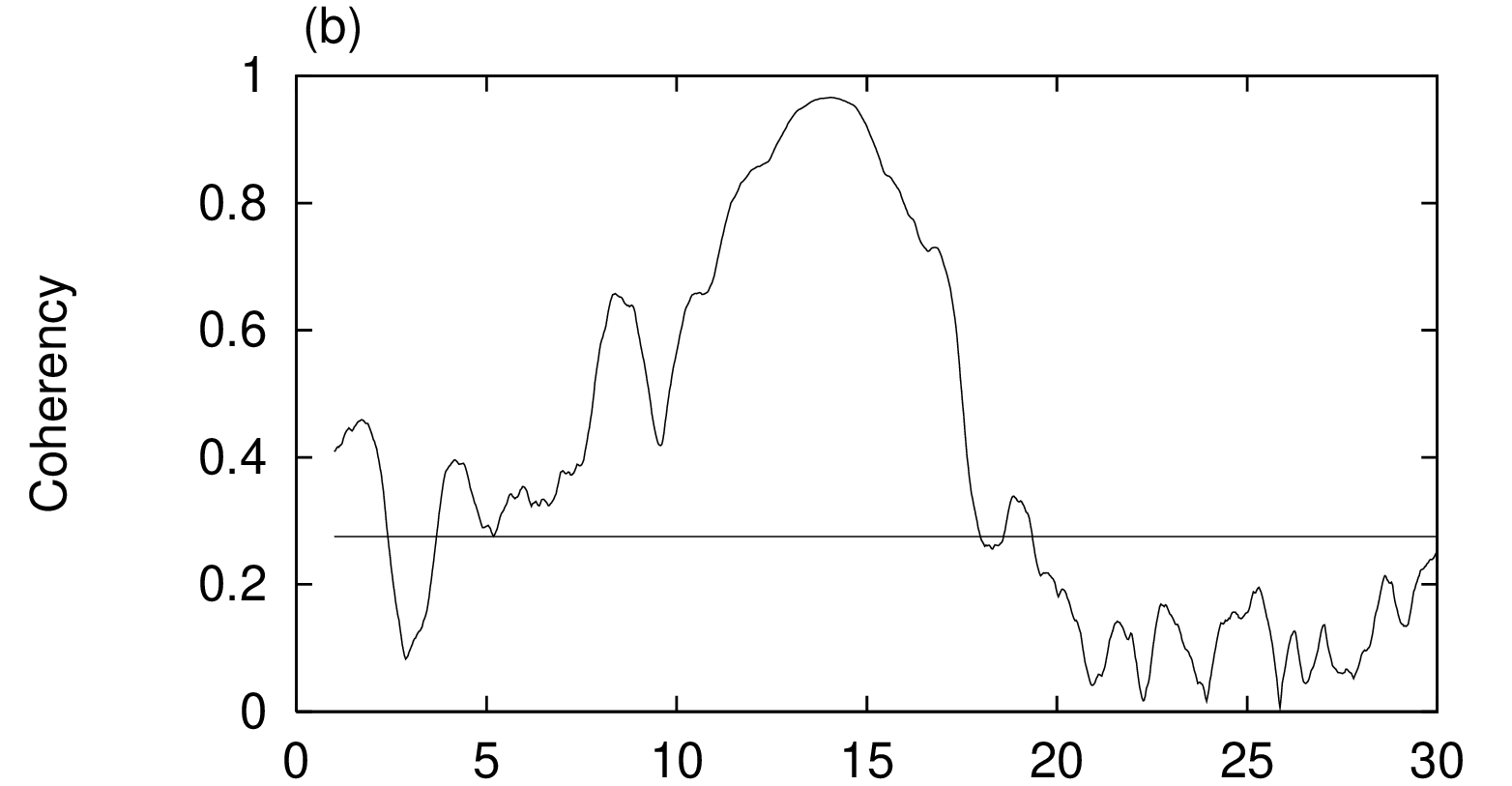}
\includegraphics[scale=0.38]{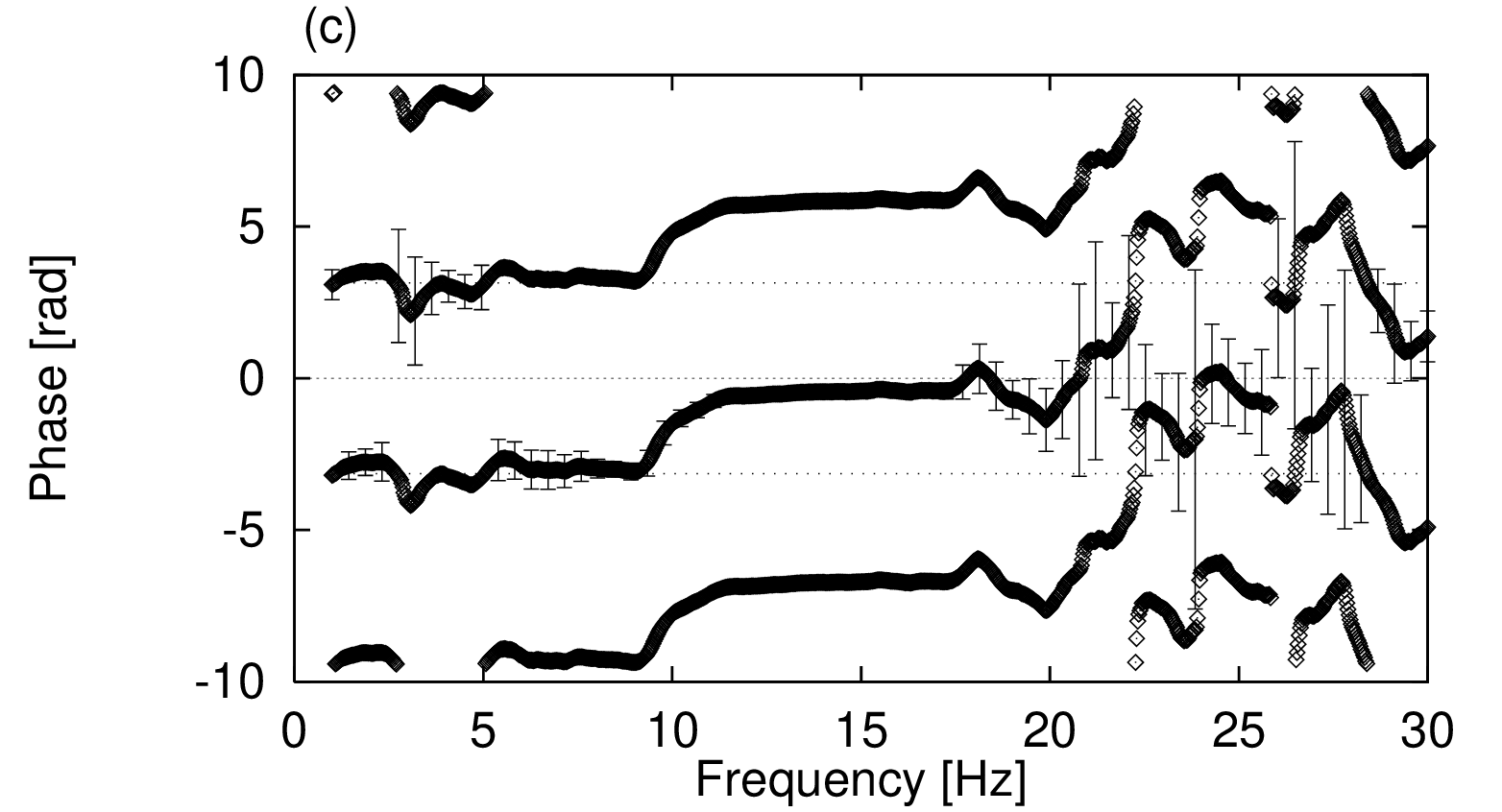}
\includegraphics[scale=.38]{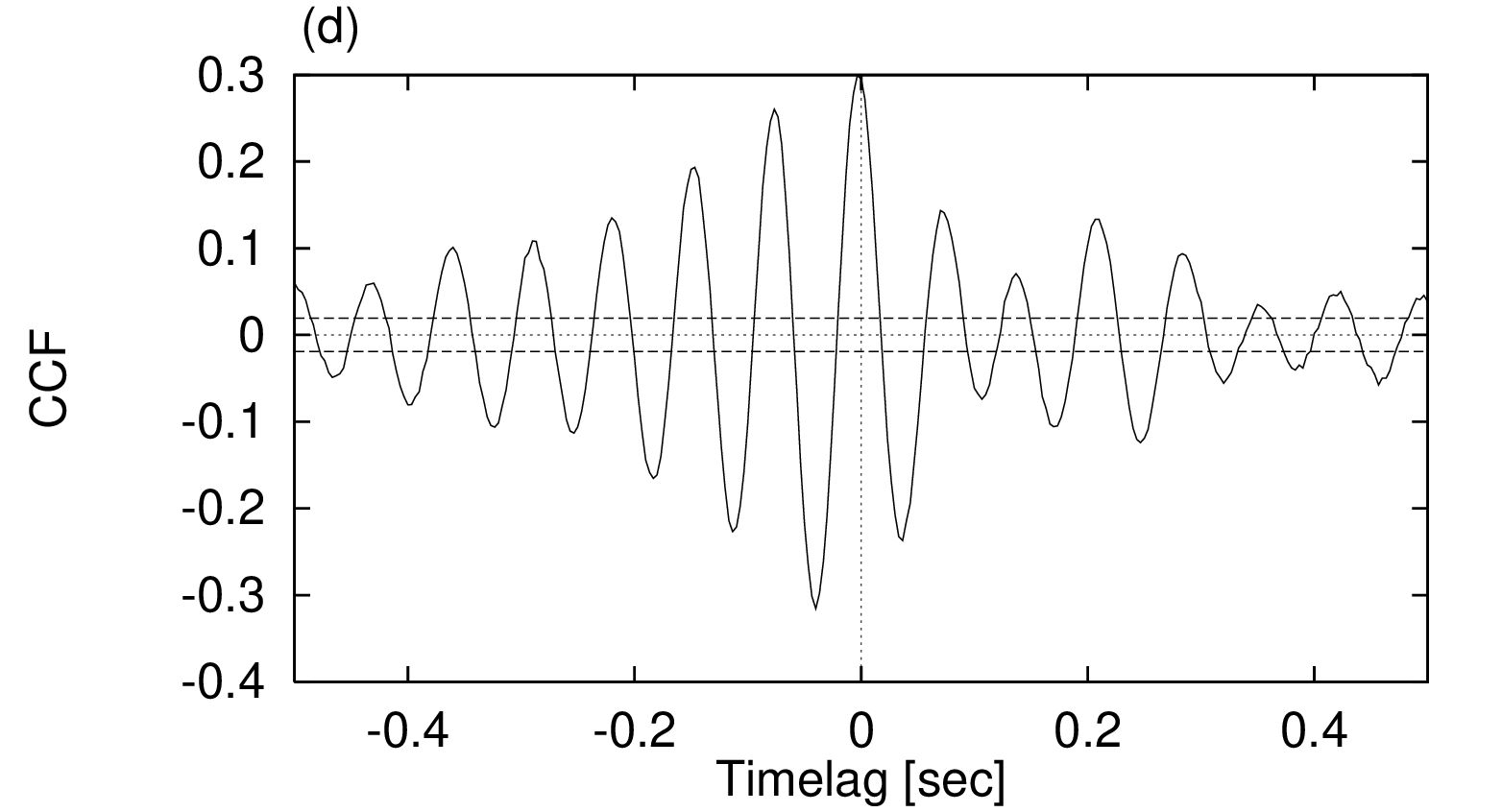}
\end{center}
\caption[]{\label{central_fig1} 
Results of a simulations study. a: power spectra (EMG: dashed
        line, ACC: solid line).  b: coherency spectrum,
        the straight line in represents the 5\% significance 
        level for the hypothesis of zero coherency, c: phase spectrum
        with 95\% confidence interval, plotted $2\pi$ periodically,
        d: cross-correlation function with 5\% significance levels for
        the hypothesis
        of zero cross-correlation assuming that at least one of the
        processes is white noise.
        The peak frequency of the EMG is higher than the resonance
        frequency of the hand. The step in the phase spectrum is determined 
        by the latter. }
\end{figure}

\begin{figure}[!ht]
\begin{center}
\includegraphics[scale=0.38]{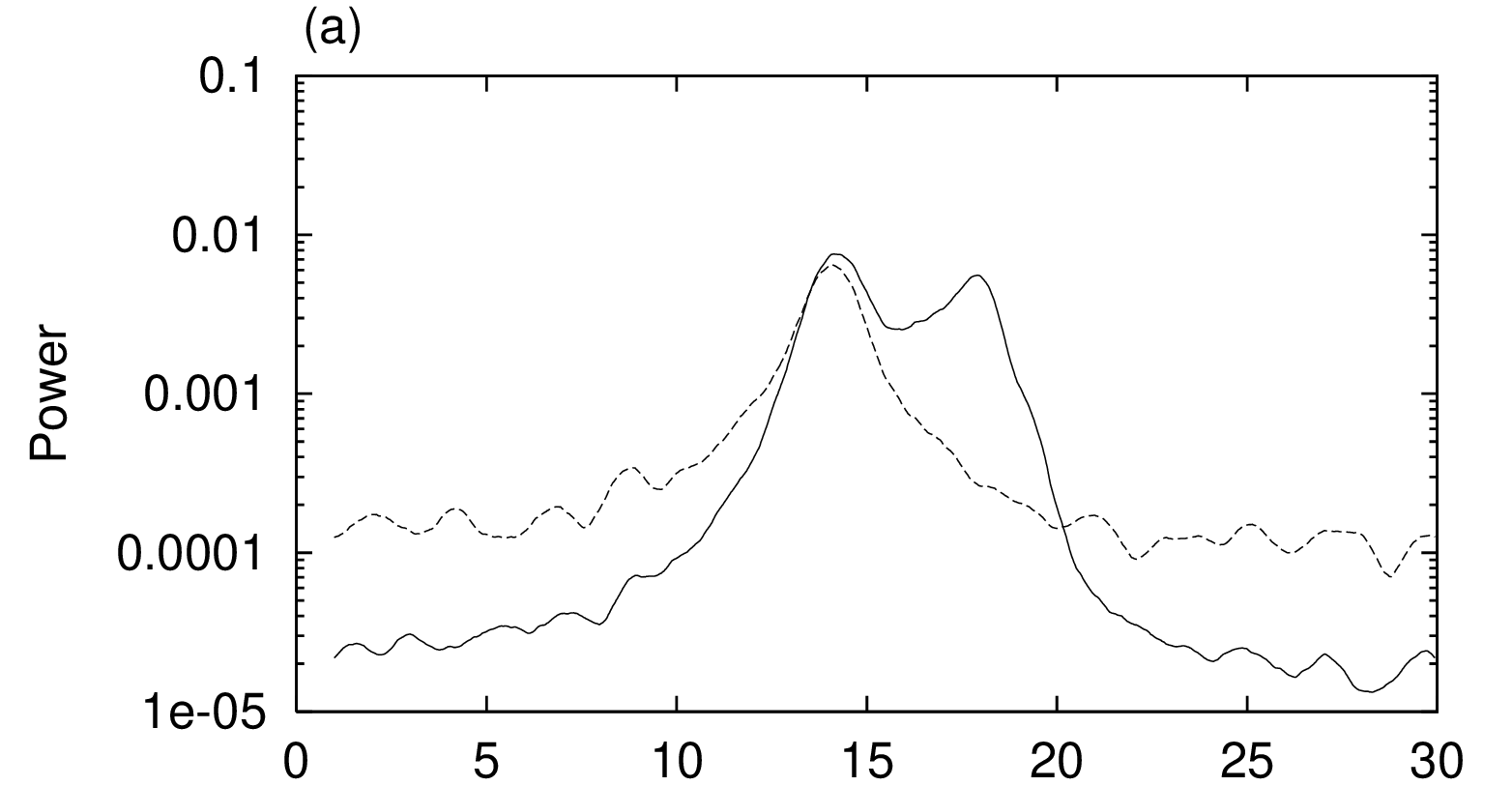}
\includegraphics[scale=0.38]{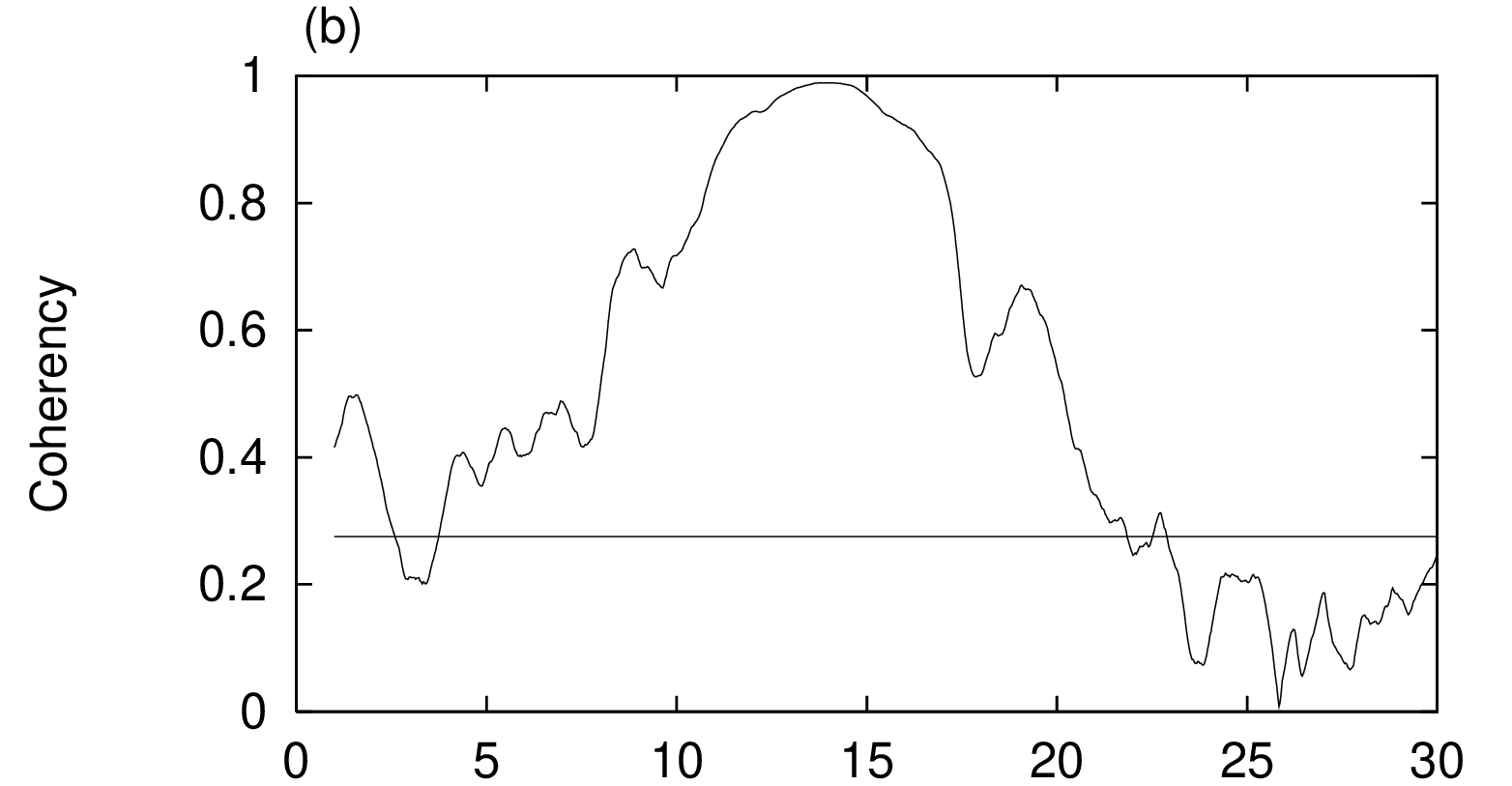}
\includegraphics[scale=0.38]{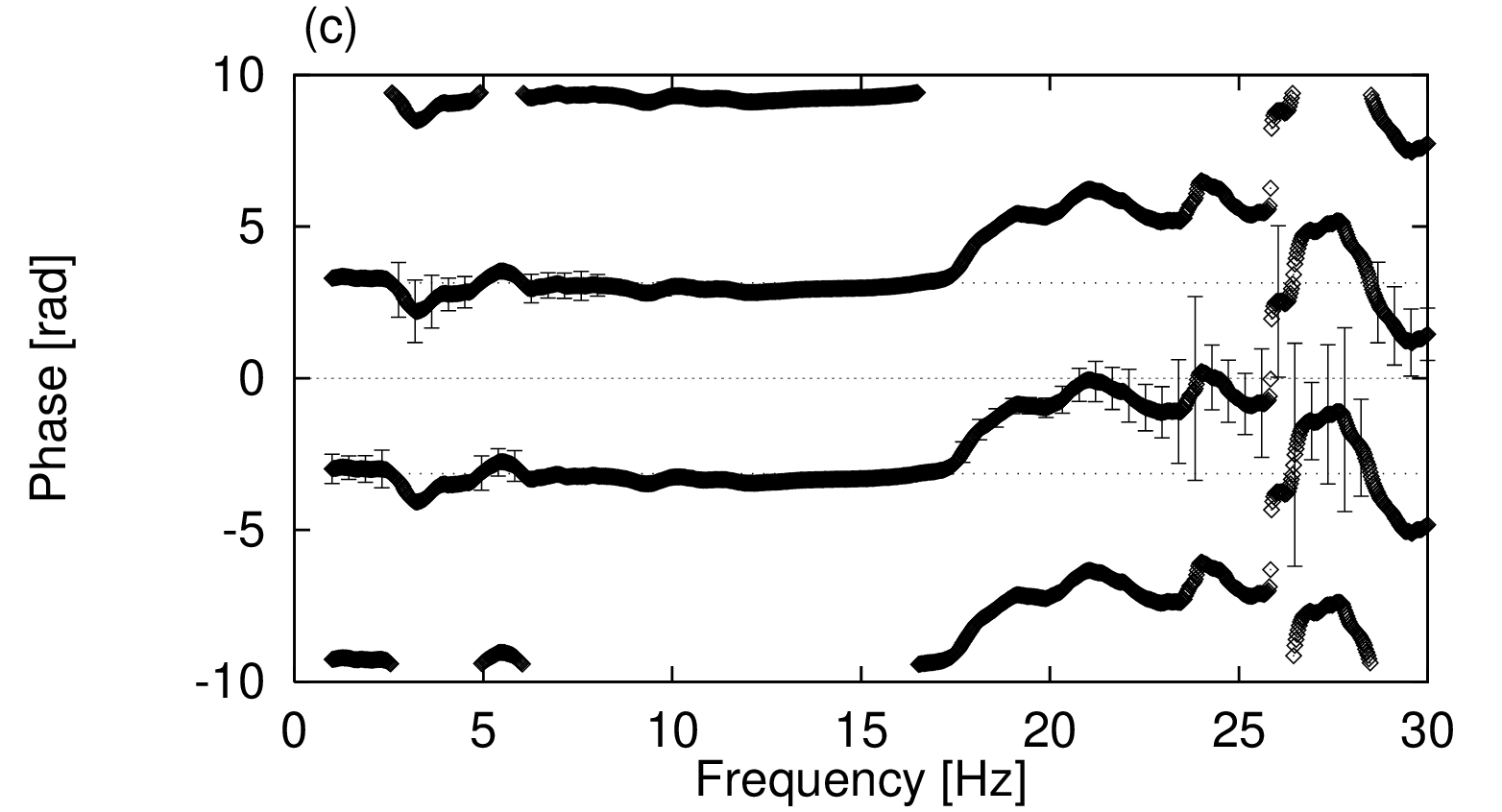}
\includegraphics[scale=.38]{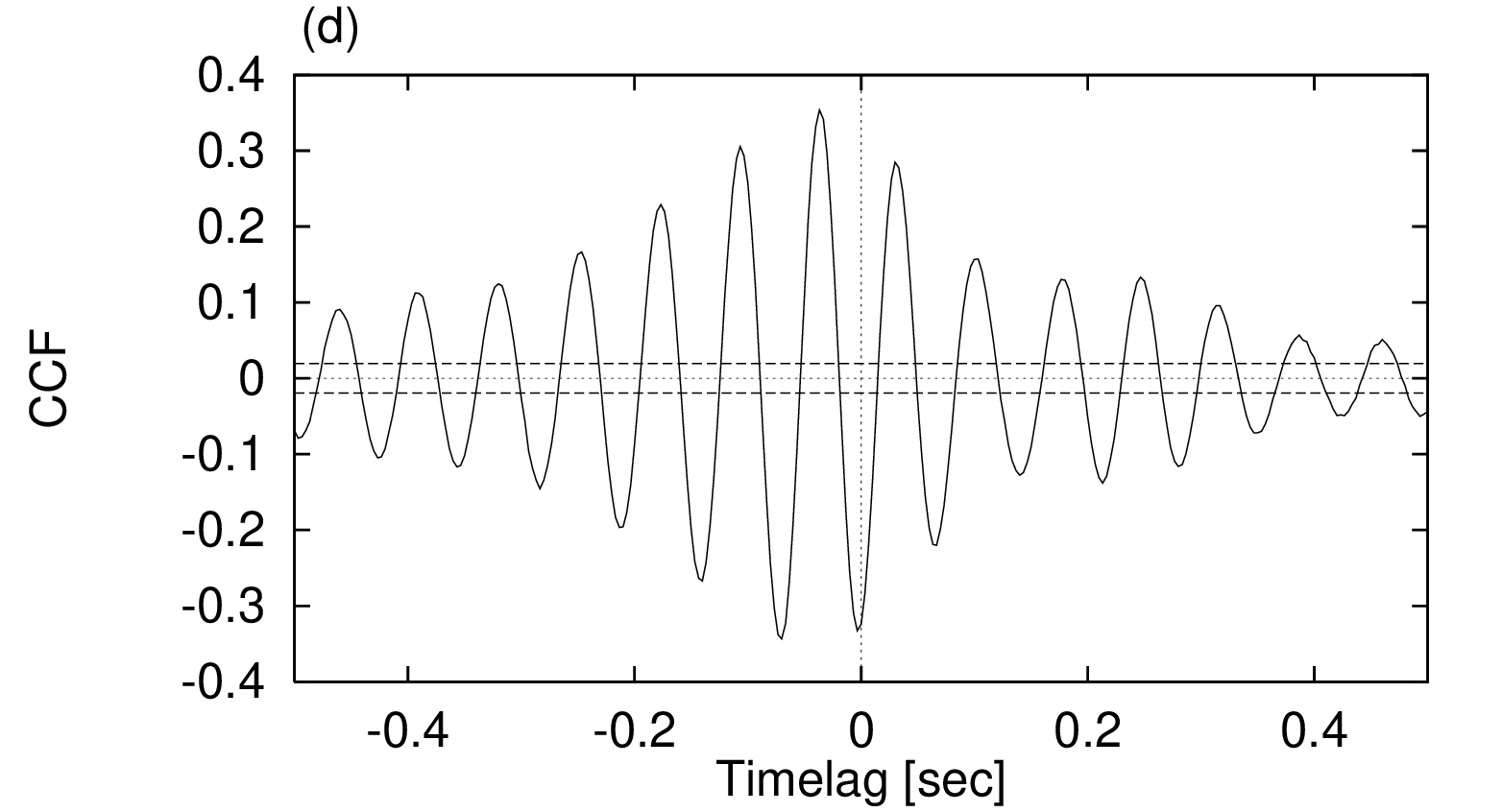}
\end{center}
\caption[]{\label{central_fig2} 
Results of a simulations study. a: power spectra (EMG: dashed
        line, ACC: solid line).  b: coherency spectrum, c: phase spectrum,
        d: cross-correlation function.
        The peak frequency of the EMG is lower than  the resonance
        frequency of the hand. The step in the phase spectrum is determined
         by the latter.}
\end{figure}

The parameters $a_1$ and $a_2$ describe the physical properties of the
hand/muscles system, i.e.~resonance frequency $1/T$ and damping rate
$\tau$. $b_1$ and $b_2$ were chosen in order to simulate the desired 
power spectra of the EMG. 
Observational noise was added to obtain the measured EMG and ACC.
The power spectrum of the ACC is given by (15) of the companion paper (Timmer et
al.~1998). The phase spectrum does not depend on the properties of the EMG 
activity. Fig.~\ref{central_fig1}
gives the results for the case that the EMG peak shows a higher
frequency (14 Hz, $b_1=1.8924, b_2=-0.9769$) than the resonance frequency of 10 Hz 
($a_1=1.9401$, $a_2= -0.9835$) of the hand.
This situation is similar to that 
often found in physiological tremor when the hand is loaded.
Fig.~\ref{central_fig1}a shows the power spectra of EMG and ACC, 
Fig.~\ref{central_fig1}b the coherency spectrum and
Fig.~\ref{central_fig1}c the phase spectrum which is displayed
$2 \pi $ periodically for a range of $\pm 3 \pi$.
 Fig.~\ref{central_fig1}d the cross-correlation function.
In Fig.~\ref{central_fig2}, results for the opposite peak frequency
 location ($a_1=1.8318, a_2=-0.9704, b_1=1.8924, b_2=-0.9769$ ) are
displayed. This situation is similar
to the finger tremor with a resonance frequency of about 20 Hz,
whereas the EMG spectrum remains usually unchanged.
Note, that the phase spectrum only depends on the resonant system.

Phase spectra like those in Fig.~\ref{central_fig1} und 
\ref{central_fig2} have been interpreted in earlier publications
only at the frequency $ \omega_0 $ of maximum coherency in terms of a
delay $ \Delta t $
(Pashda and Stein 1973; Elble and Randall 1976) by
 $ \Delta t = \Phi(\omega_0) / \omega_0 $. Obviously, such
an interpretation might be misleading, since no delay is included
in the process. The frequency of maximum coherency is determined by
the frequency dependent signal to noise ratio ((27) of the companion
paper) und usually occurs at
the peak frequency of the EMG. The phase spectrum evaluated
at the peak frequency of the EMG only contains the information
whether the resonance frequency of the hand/muscle system is located 
above or below the EMG peak frequency.

The cross-correlation function is difficult to interpret. 
The period of the cross-correlation function is mainly determined by the 
frequency of highest coherency and modulated by frequencies of
medium coherency.
Obviously, the time-lag where the cross-correlation function shows its maximum
may, in general, not be interpreted as a time delay between the processes. 
The magnitude of the cross-correlation function is similar for
positive and negative lags. Especially for negative time lags, it
 exceeds significantly the 5 \% significance
level of $ \pm 1.96 N^{-1} $ of zero cross-correlation 
derived under the assumption that at least one of the independent
processes is white noise. The reason is that the autocorrelation function
of both processes and the true cross-correlation function
enters the correlation structure of the estimated cross-correlation
function here. Thus, large values of the cross-correlation function for
negative time lags, suggesting an effect from the ACC on the EMG,
do not necessarily indicate the involvement of a reflex loop.

\subsection {Application to measured data and modeling}
\begin{figure}[!h]
\begin{center}
\includegraphics[scale=0.38]{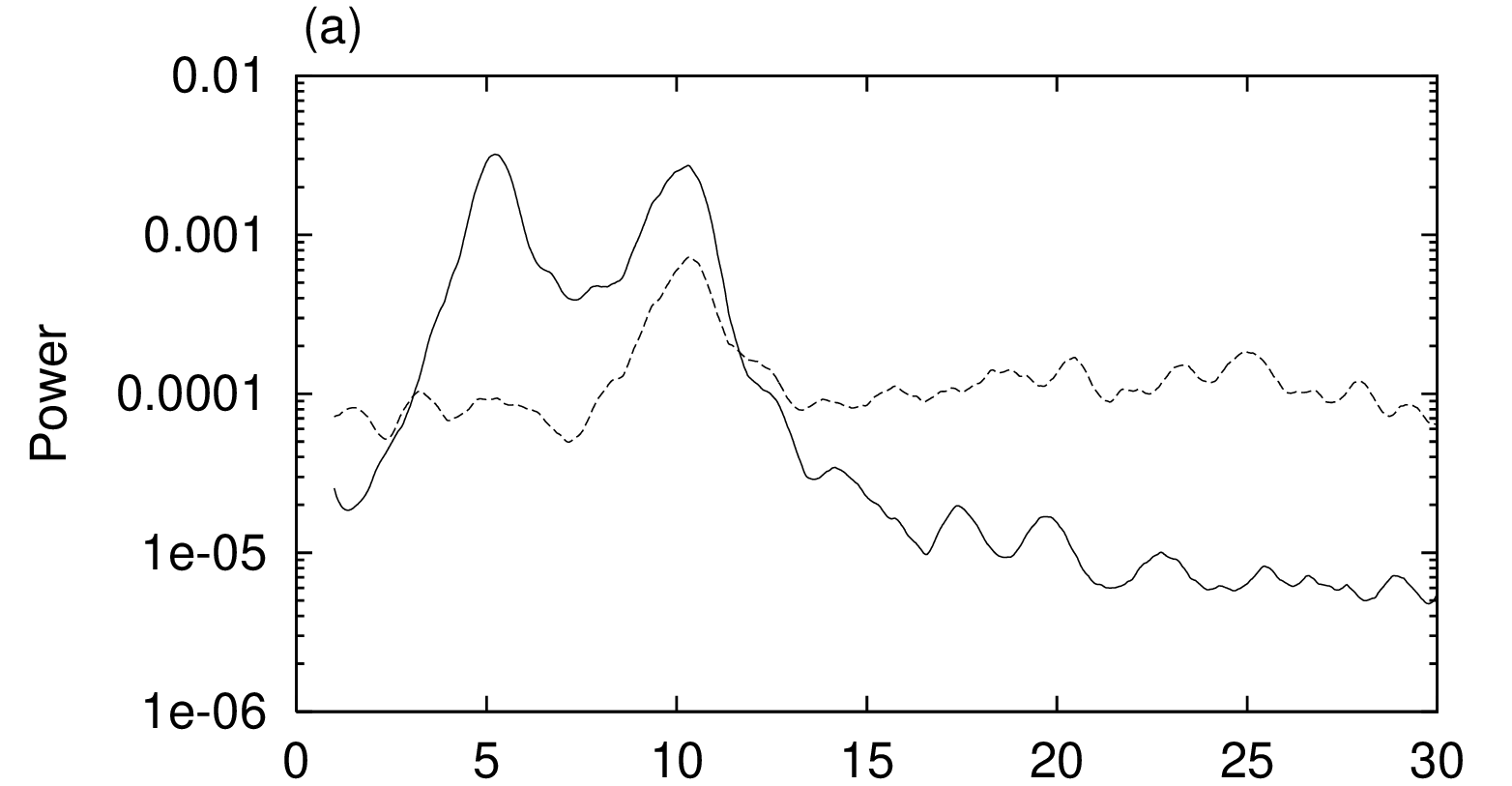}
\includegraphics[scale=.38]{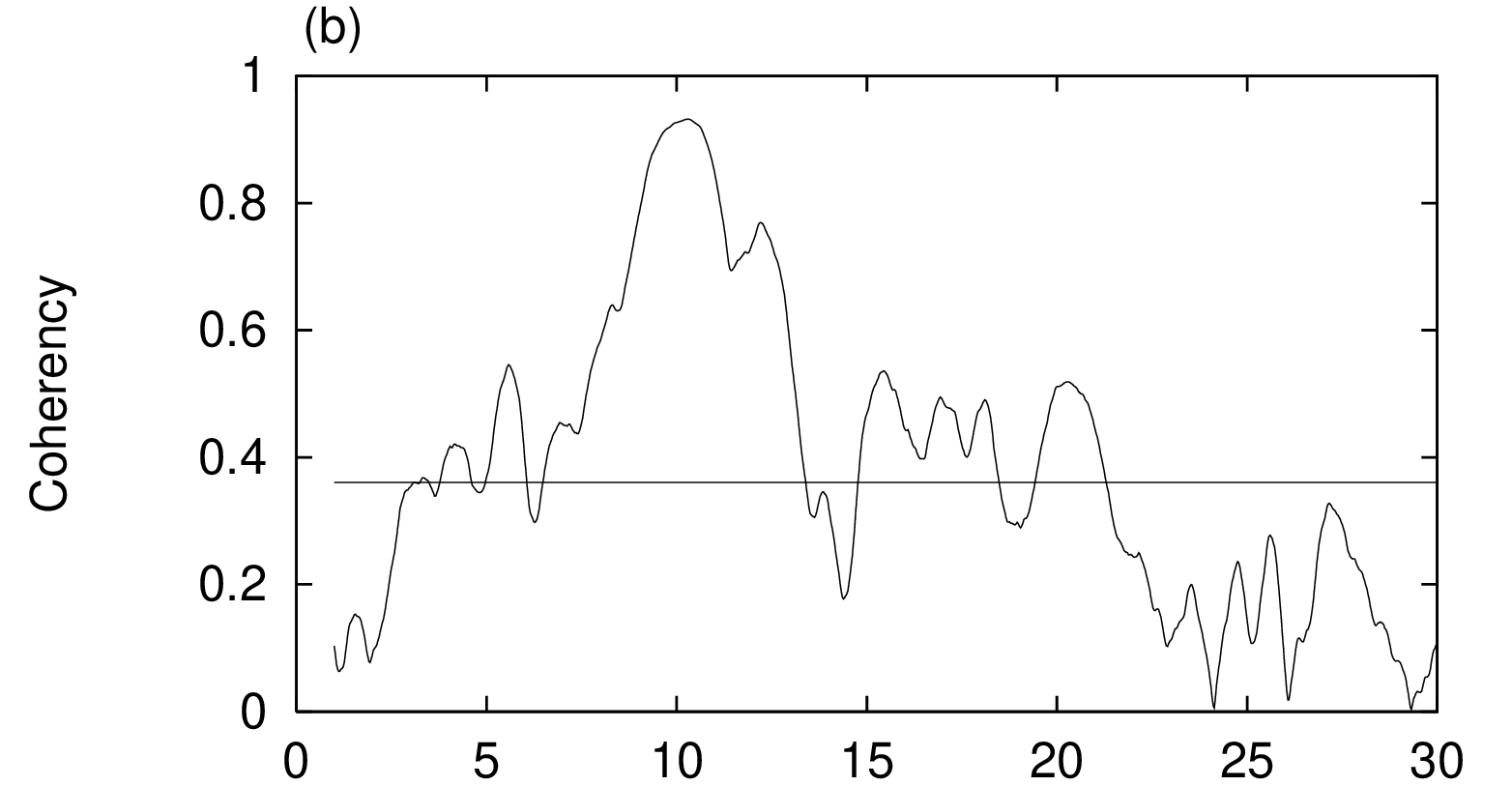}
\includegraphics[scale=.38]{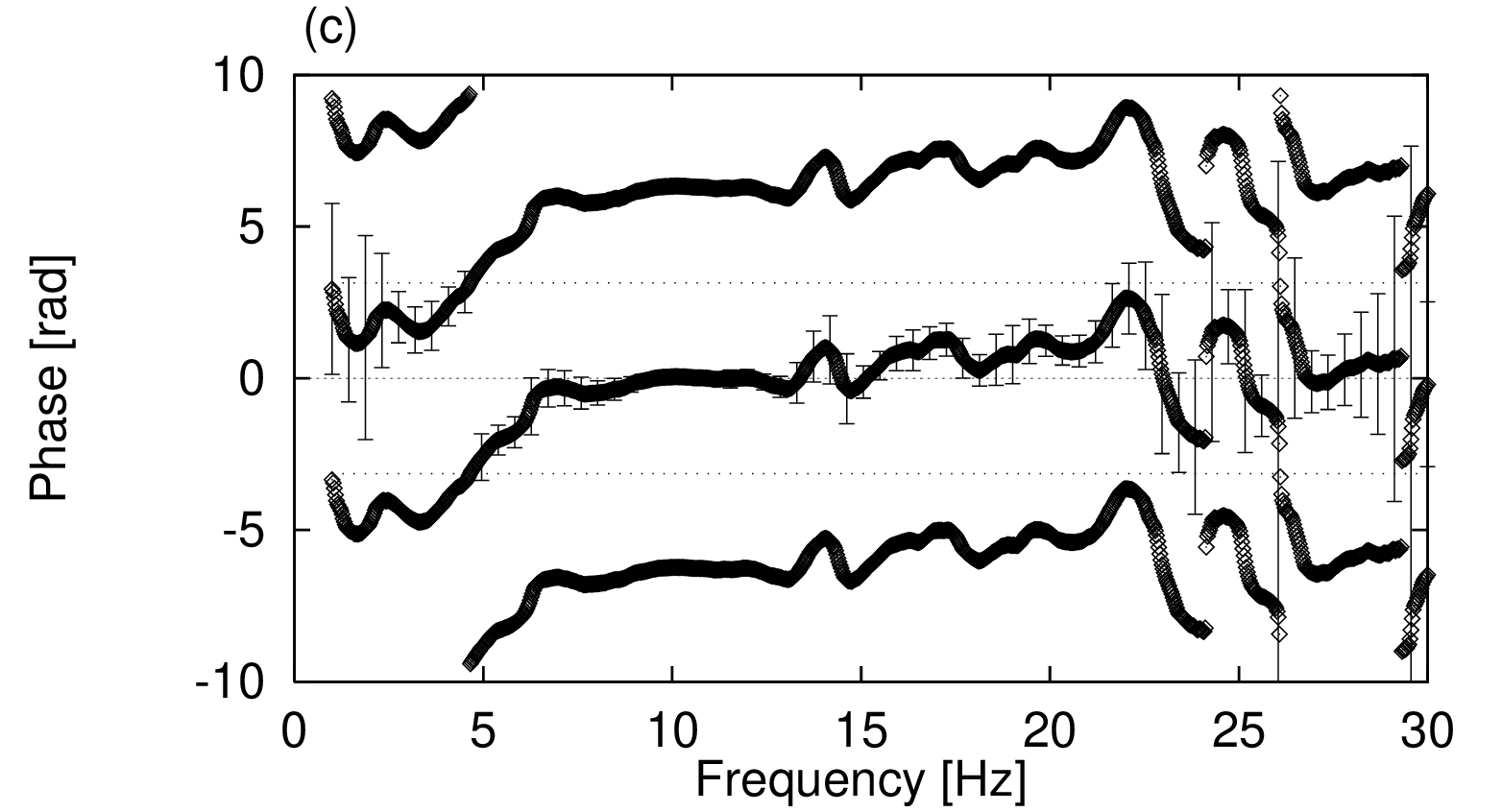}
\includegraphics[scale=.38]{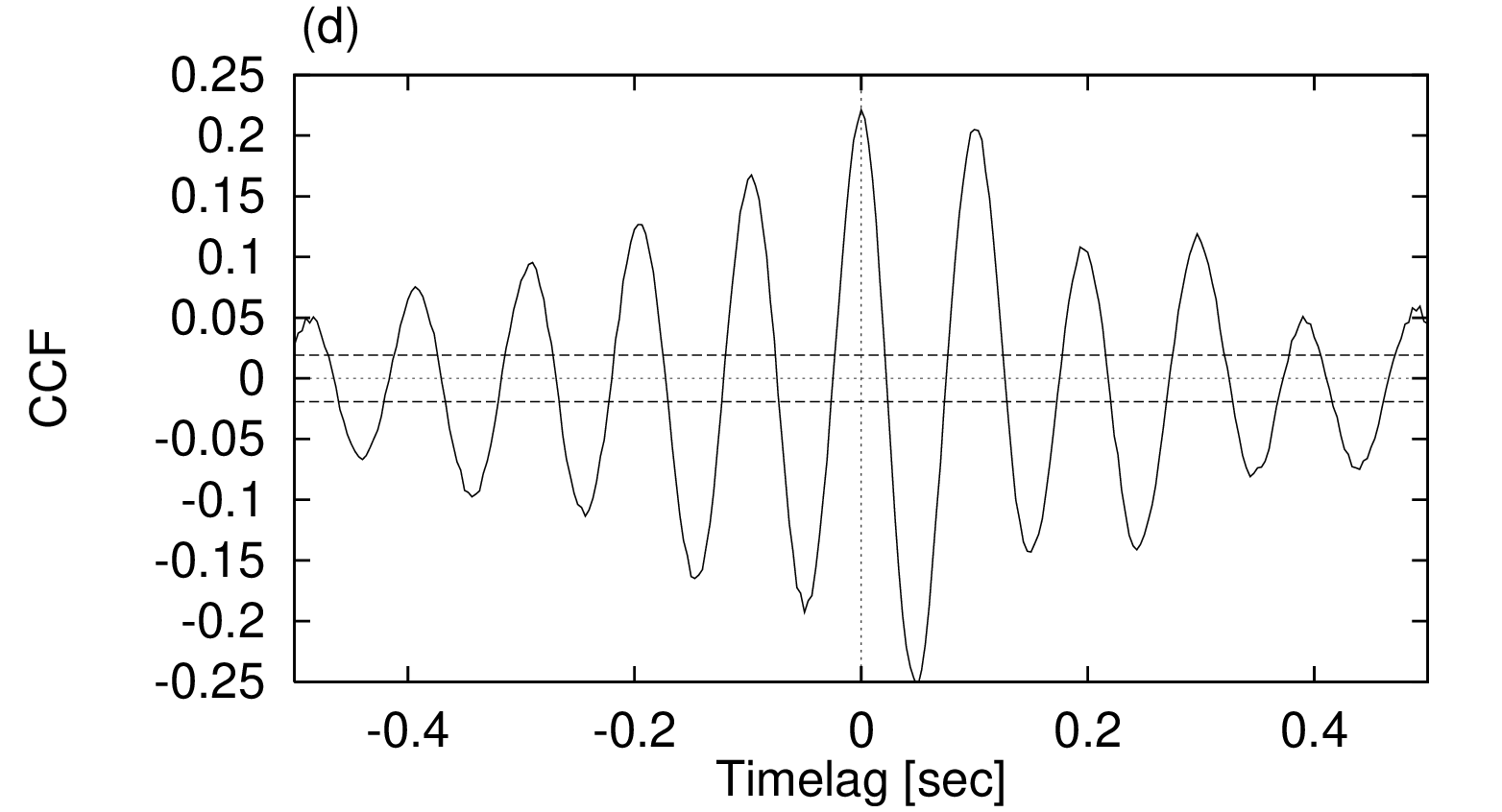}
\end{center}
\caption[]{\label{real_data_non_reflex} 
Physiological tremor with EMG synchronization.
        a: power spectra (EMG: dashed line, ACC: solid line), b: coherency 
        spectrum, c: phase spectrum, d: cross-correlation function.}
\end{figure}

\begin{figure}[!h]
\begin{center}
\includegraphics[scale=0.38]{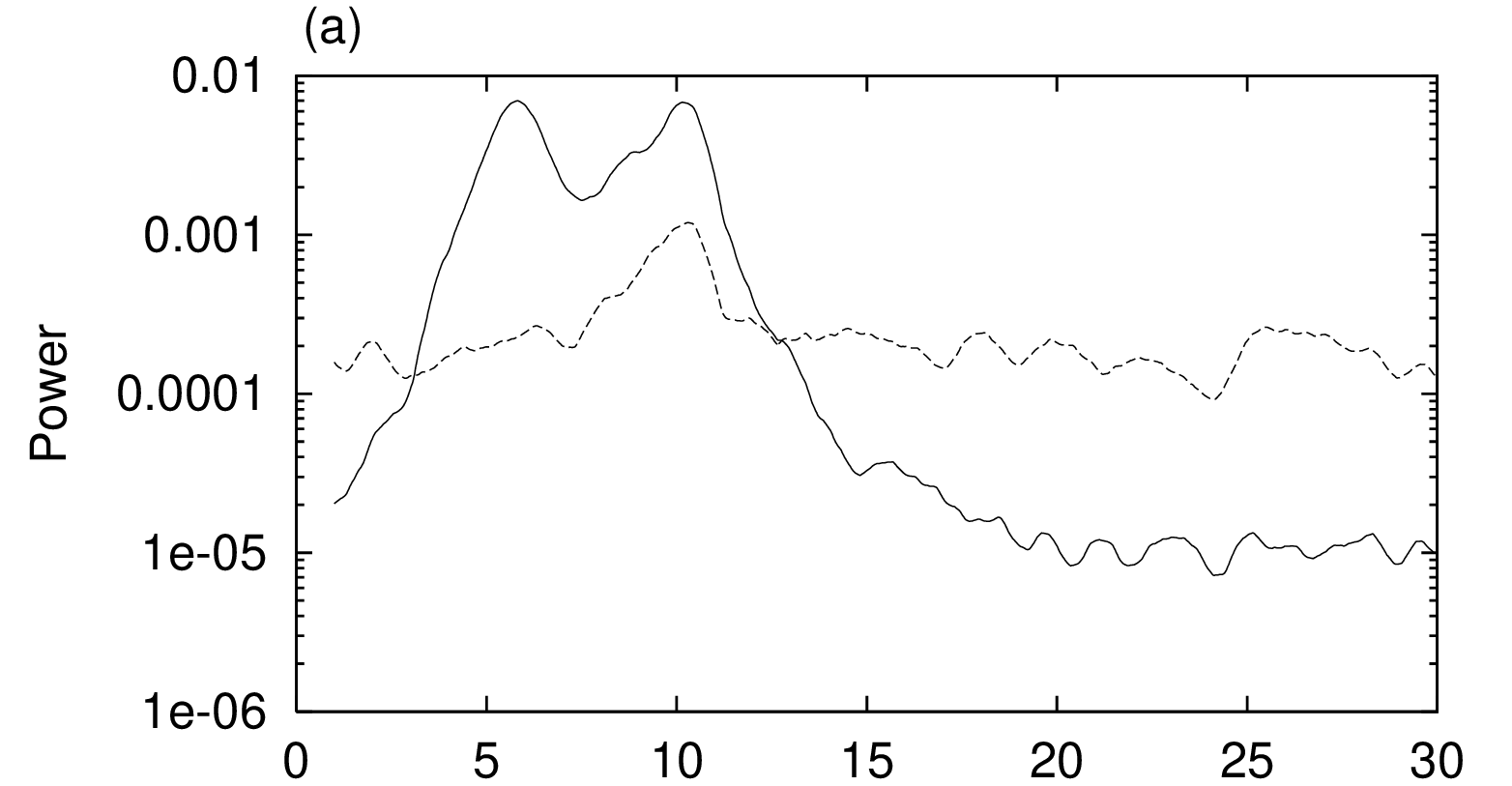}
\includegraphics[scale=.38]{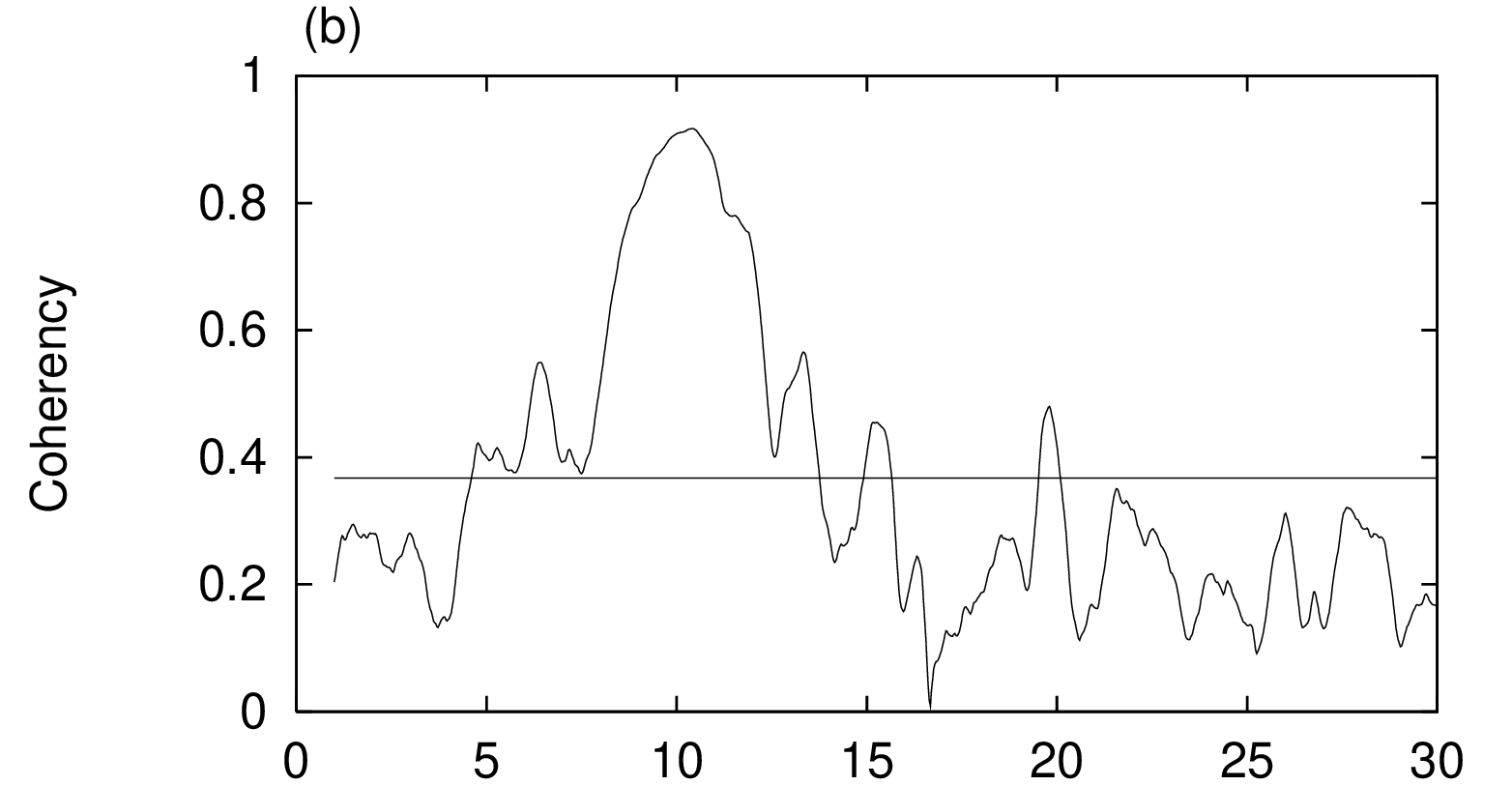}
\includegraphics[scale=.38]{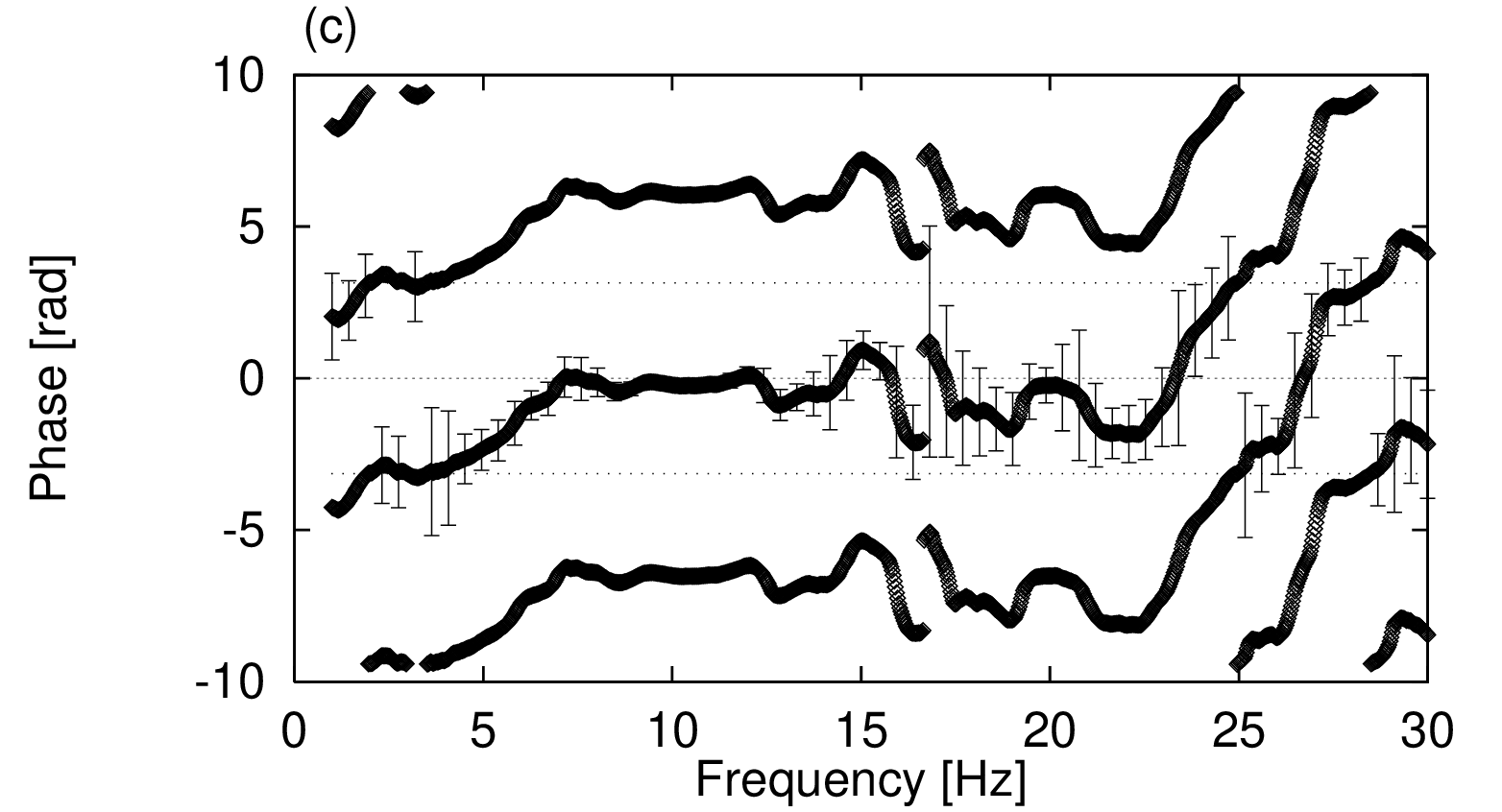}
\includegraphics[scale=.38]{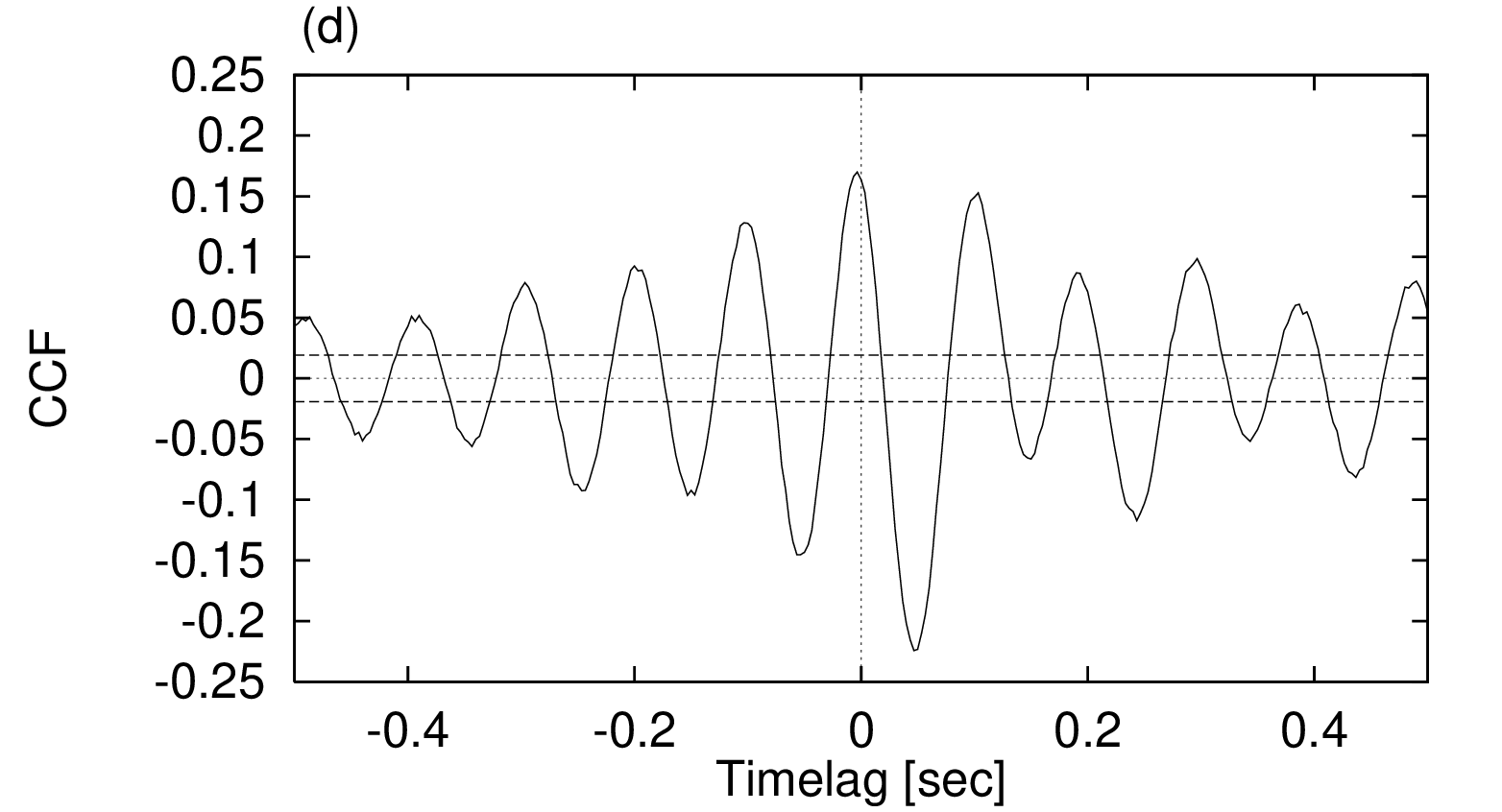}
\end{center}
\caption[]{\label{simu_data_non_reflex} 
Realization of the model fitted to the data of Fig.3.
        a: power spectra (EMG: dashed line, ACC: solid
        line), b: coherency spectrum,
        c: phase spectrum, d: cross-correlation function.}
\end{figure}

Fig.~\ref{real_data_non_reflex} shows the results of an analysis of
a physiological hand tremor showing an EMG synchronization at approximately
$10.5$~Hz. 
It resembles to Fig.~\ref{central_fig1}. The behavior of the 
coherency can be explained as an effect of observational noise 
by (27) of the companion paper. 
In order to show that these results can be interpreted in the framework
of linear models we fitted a model of seven parameters 
according to (\ref{nr1},\ref{nr2}) to
the empirical data. The seven parameters are 
periods and relaxation times describing the EMG and the mechanical
properties of the hand, the variance of the observational noise
covering EMG and ACC and the amount of additional unsynchronized EMG
activity. By Equations (2,3,4,5,13,15,22,23,27) of the companion paper
these parameters can be estimated from the data. 
Fig.~\ref{simu_data_non_reflex} shows the results of a simulation of
the fitted model. All quantitative details of Fig.~\ref{real_data_non_reflex}
are reproduced by the model.

\section {Reflex mechanisms} \label{reflexes}
Pioneered by the work of Lippold
(Lippold 1957, 1970) various experiments were performed to clarify the 
effect of reflex mechanisms on physiological tremor. In most of these
experiments a mechanical or
electrical stimulus was applied to measure the response of the
muscle (Rack et al.~1978, Hagbarth and Young 1979, Matthew 1994). 
By cross-spectral analysis we studied the role of reflexes in 
physiological tremor without perturbing the system, i.e. the stimulus
for the reflexes is the tremor itself.

\subsection {Simulations}
To investigate the phenomena that might
be induced by reflex mechanisms, we use a stochastic feedback system
which applies a sigmoidal nonlinearity describing the activation function
of the motoneurons. The effects of such an activation function
in a deterministic framework simulating spinal as well as cortical
feedback delays were introduced by Stein and Oguzt\"oreli 
(Stein and Oguzt\"oreli 1975, 1978; Oguzt\"oreli and Stein 1975, 1976,
1979).
The reflex model of Stein and coworkers includes the muscle length
and velocity. We use the tremor amplitude and velocity. Because of the small
amplitude these quantities are proportional. 
In order to study the influences of reflex mechanisms 
on the spectra, we used the following model. 
Denoting the reflex induced EMG by $ y(t) $,
the unsynchronized EMG activity by $ \nu(t) $,
the movement of the hand by $ x(t) $, a possible time
delay of the effect of the EMG on the hand by $ \Delta t $, and 
the reflex loop delay by $\delta t$ we receive:
\begin{eqnarray} 
 y(t) & = & \kappa f(x(t-\delta t) , \dot{x}(t-\delta t)) 
        \label{reflexmodel1} \\
 x(t) & = & a_1 x(t-1) + a_2 x(t-2) \\\nonumber
        && \qquad\qquad\qquad   + \nu (t-\Delta t) + y(t-\Delta t)
        \label{reflexmodel2}  \quad .
\end{eqnarray}
 For the function $ f(.) $ we chose the tangens hyperbolicus
with gain $ \kappa $; $a_1$ and $a_2$ describe the physical properties
of the musculosceletal system. $x(t-\delta t)$ models the length dependent 
and $\dot{x}(t-\delta t)$ the velocity dependent reflex.
Observational noise was added to both processes. Its standard
deviation was 10 \% of the standard deviation of the noise free data.
Note, that (\ref{reflexmodel1},\ref{reflexmodel2}) form
a nonlinear stochastic delay differential equation.

\begin{figure}[!h]
\begin{center}
\includegraphics[scale=0.43]{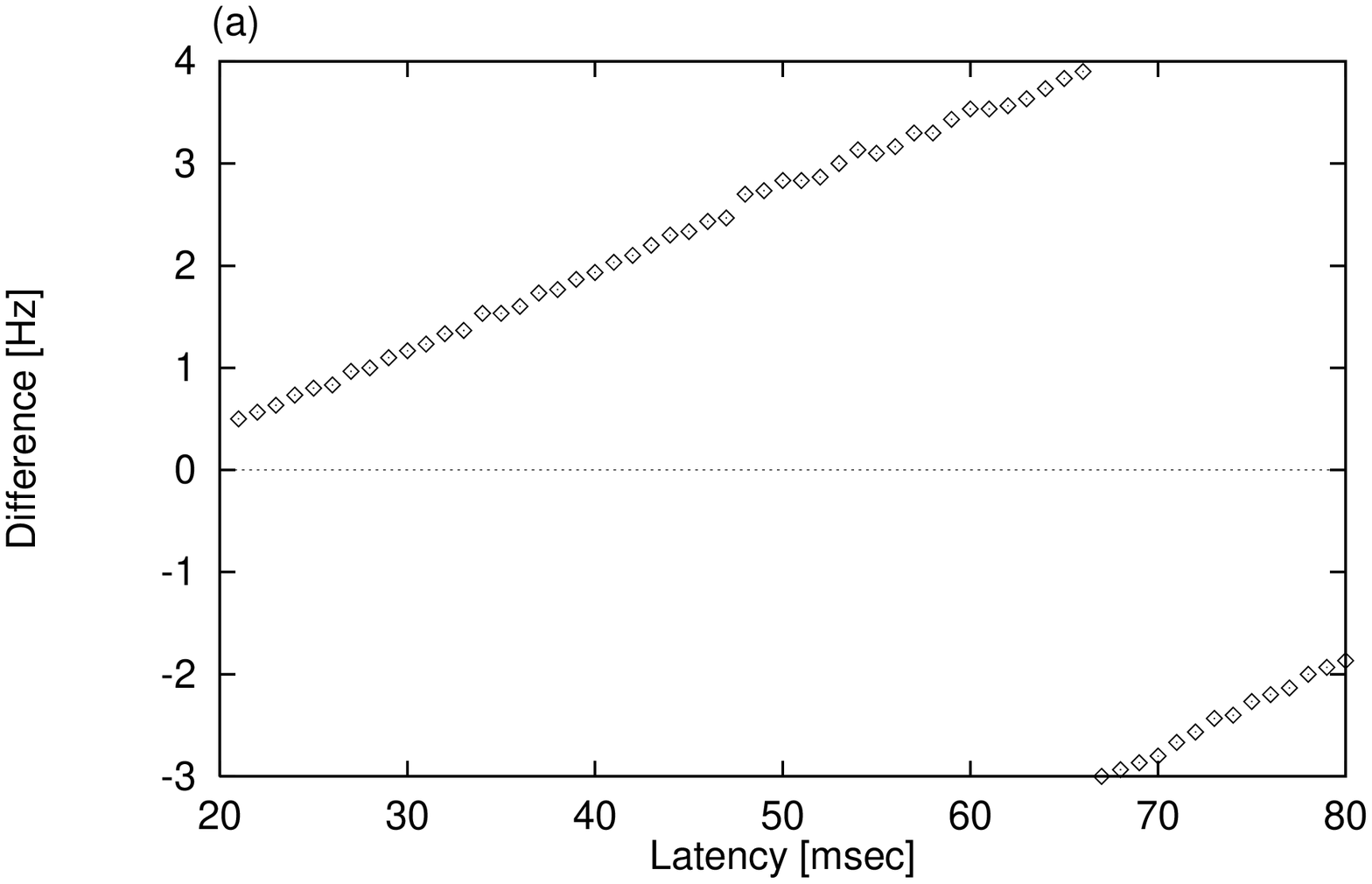}
\includegraphics[scale=0.43]{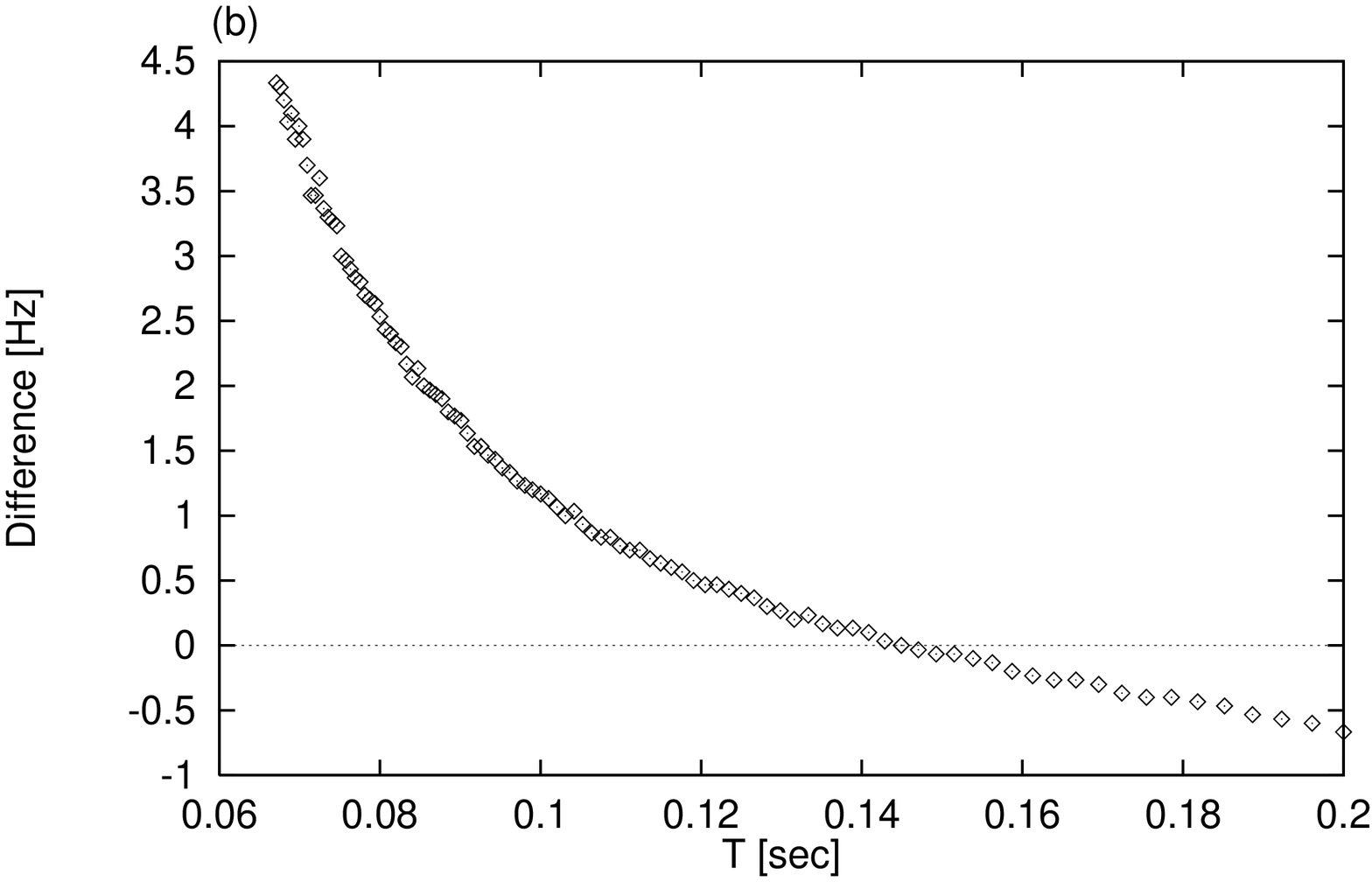}
\end{center}
\caption[]{\label{T_von_reflex} 
Change of peak frequency due to the reflex.
        a: Dependence on the delay of the reflex loop for a fixed
        characteristic
        period (100 ms) of the mechanical system, b: Dependence on
        the period of the mechanical system for a fixed delay of the
        reflex loop of 30 ms.}
\end{figure}

Due to the nonlinear activation function, in general,
the spectra can not be calculated analytically.
We performed simulation studies in order to investigate the 
behavior of the spectra. 
The unique result for various settings of the parameters is, that 
\begin {itemize}
\item the EMG power spectrum shows a peak at the same frequency as the ACC
      power spectrum. The former might be hidden by the observational noise.
\item in accordance with Oguzt\"oreli and Stein (1975) the  
      frequency of the oscillation is a complex function 
      of the delays and the properties of the hand and muscles.
        Fig.~\ref{T_von_reflex}a displays the reflex induced frequency 
        shift in dependence on the delay of the reflex loop. 
        Fig.~\ref{T_von_reflex}b shows this shift as a function of
        the characteristic period of the mechanical system.
\item the phase spectrum is given by the mechanical properties of the 
        hand/muscle system represented by $a_1$ and $a_2$, i.e.~not dependent
      on the reflexes.
\item  the peaks of the ACC are sharper in presence of a reflex than 
        without, in accordance with experimental findings (Marsden 1984). 
\end{itemize}

In general a non-reflex and a reflex component might show up
in the EMG. For this case, the models (\ref{nr1},\ref{nr2}) and 
(\ref{reflexmodel1},\ref{reflexmodel2}) have to be combined to:
\begin{eqnarray}
  y(t) & = & \kappa f(x(t-\delta t) ,\dot{x}(t-\delta t)) \\\nonumber
       && \qquad\qquad\qquad +  b_1 y(t-1) + b_2 y(t-2) + \epsilon(t) \label{reflex3} \\
x(t) & = & a_1 x(t-1) + a_2 x(t-2)  \\\nonumber 
        && \qquad\qquad\qquad   + \nu(t-\Delta t) + y(t-\Delta t) \label{reflex4}  \quad .
\end{eqnarray}

\begin{figure}
\begin{center}
\includegraphics[scale=0.36]{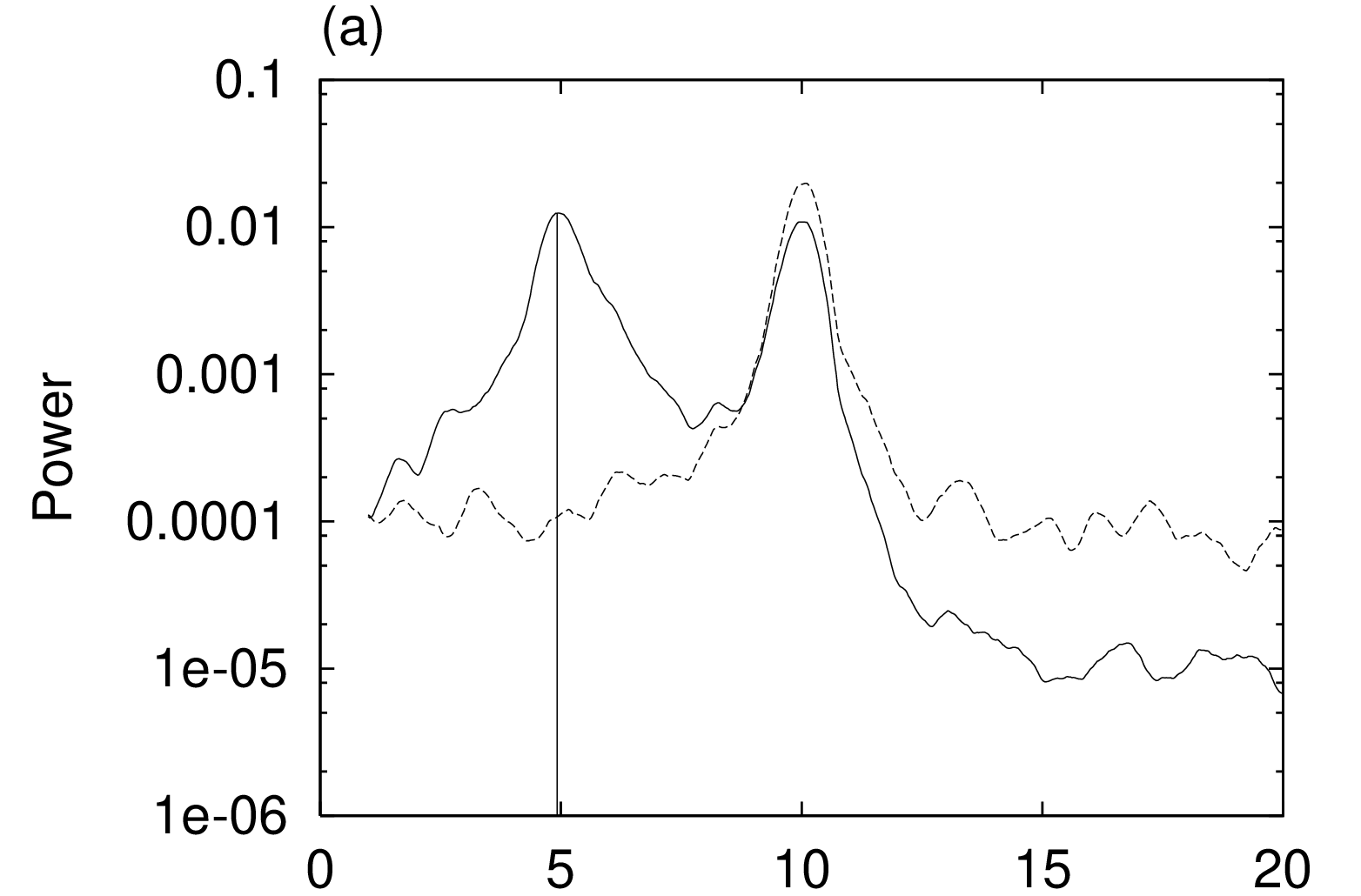}
\includegraphics[scale=0.36]{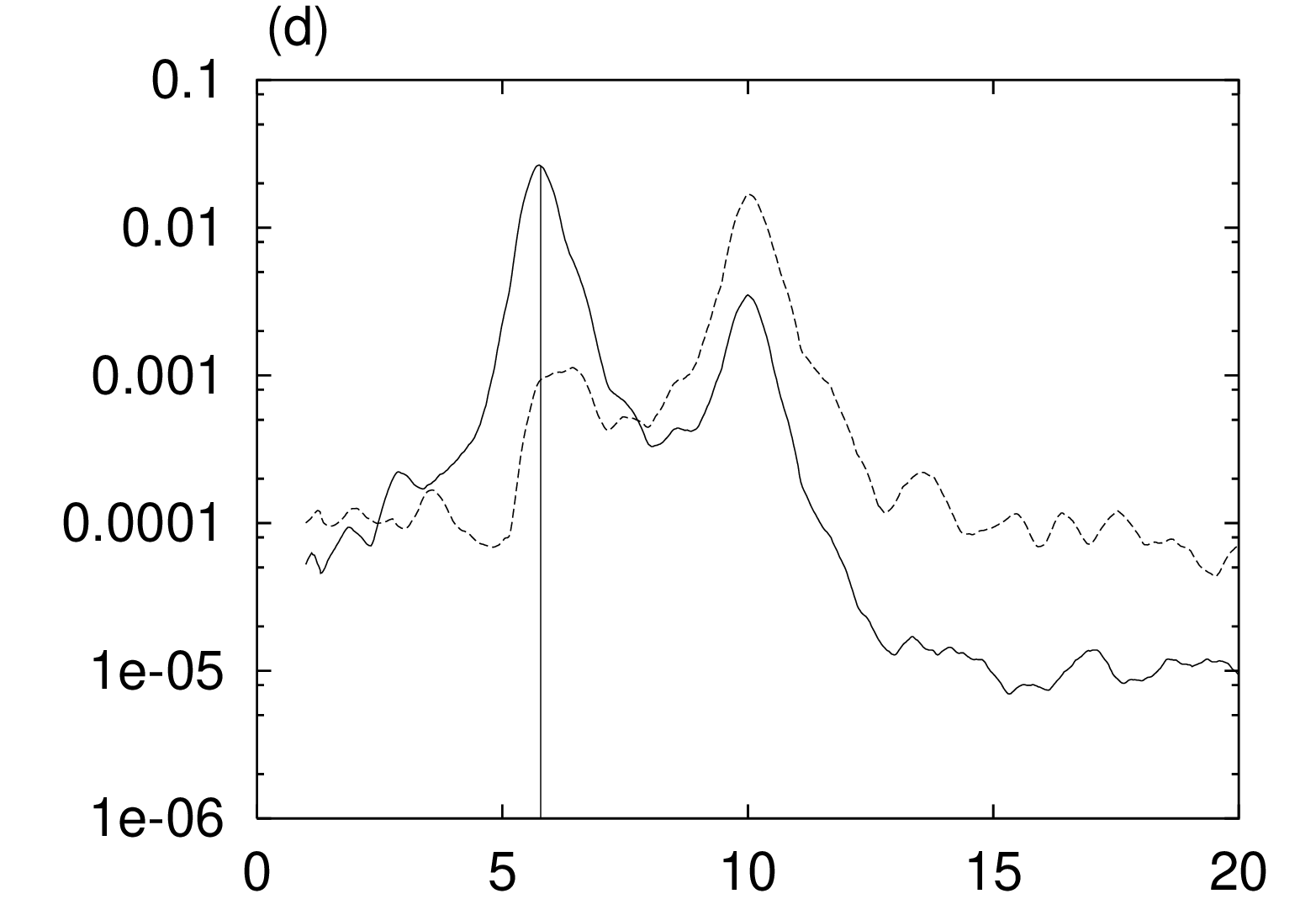}
\includegraphics[scale=0.36]{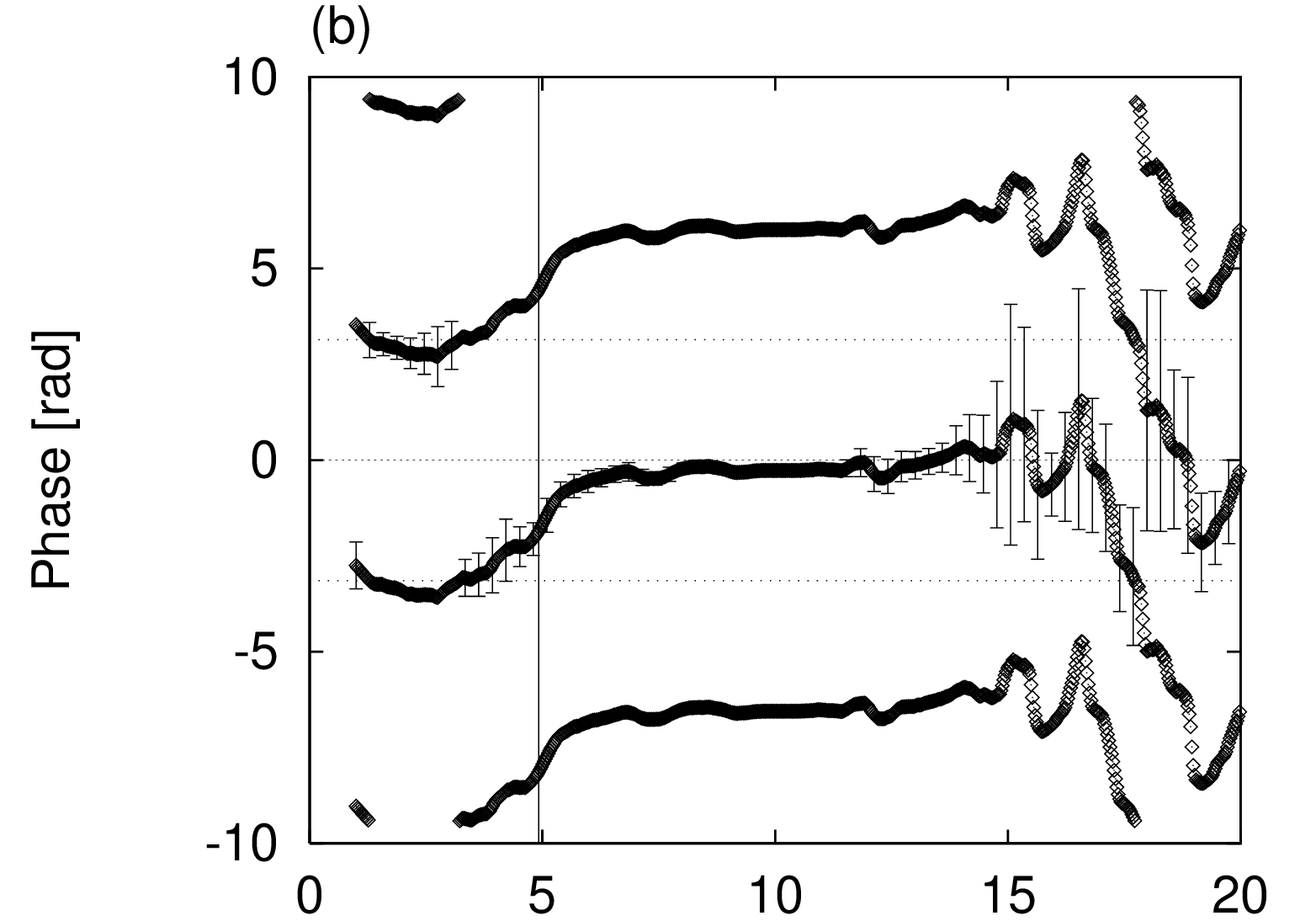}
\includegraphics[scale=0.36]{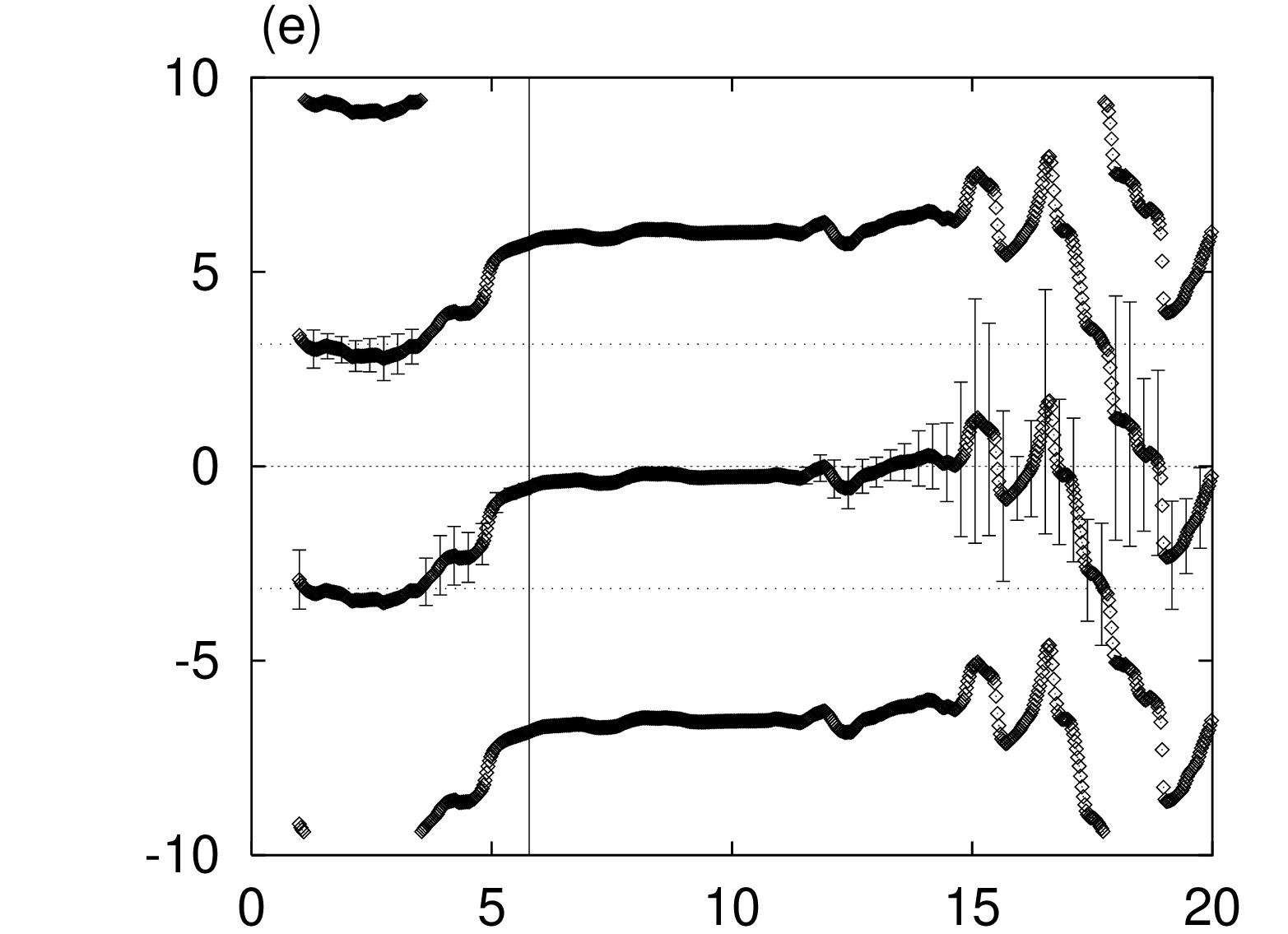}
\includegraphics[scale=0.36]{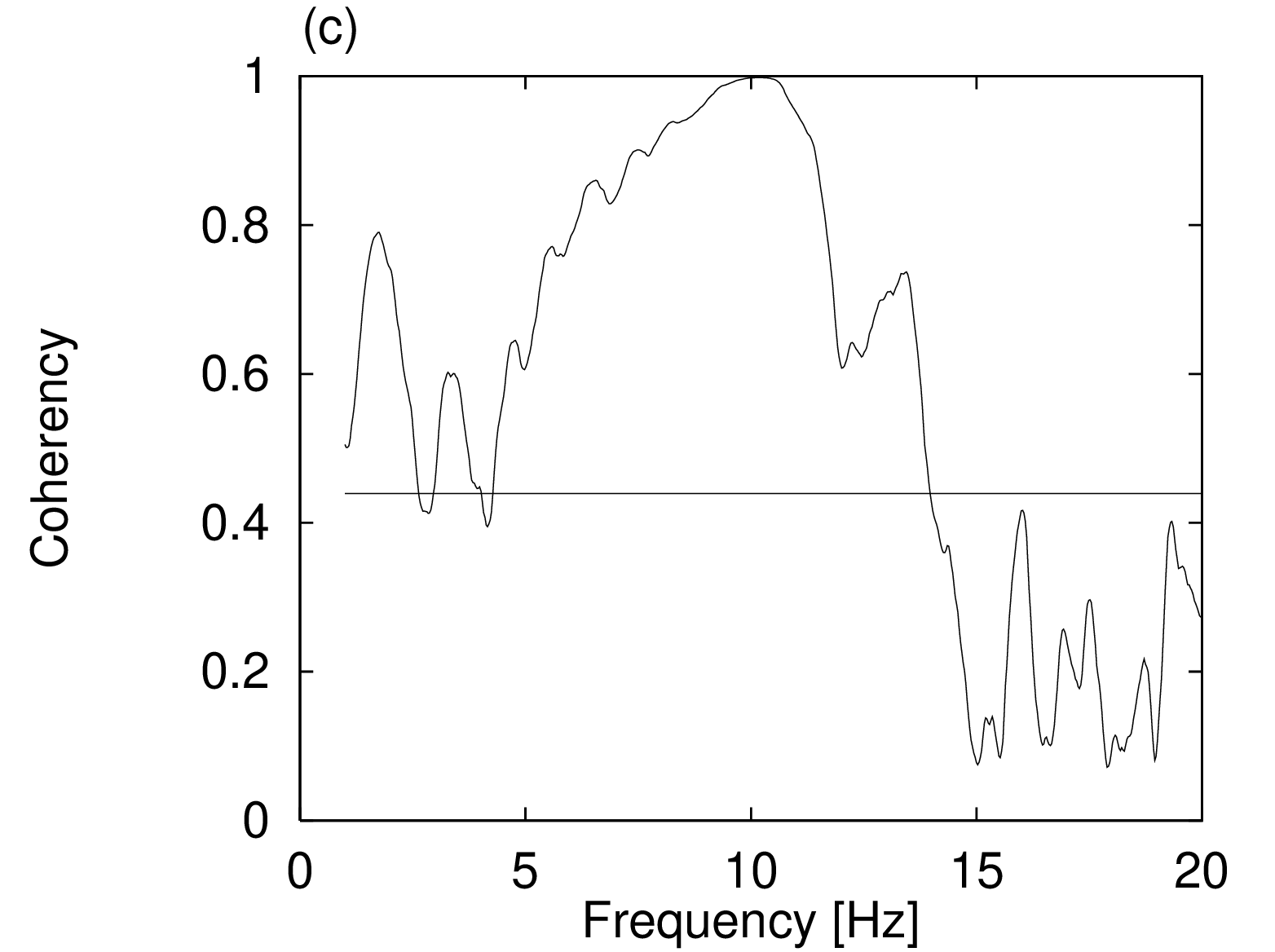}
\includegraphics[scale=0.36]{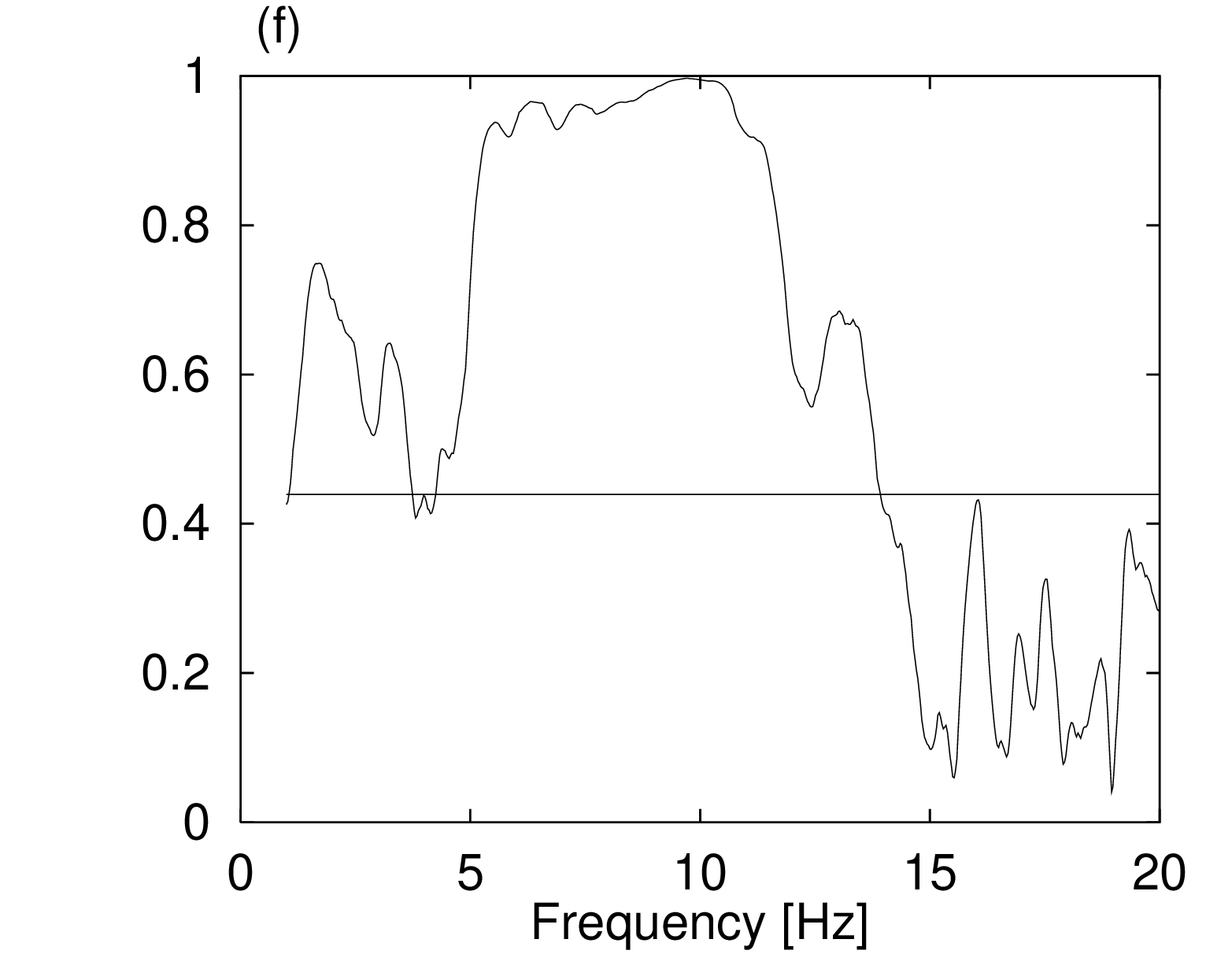}
\end{center}
\caption[]{\label{reflex_phase_fig2} 
Results of a simulation study. a: EMG (dashed) and ACC
          (solid) power spectra in
        the case of no reflex. b: corresponding phase spectrum c: 
        coherency spectrum. d: EMG and ACC power spectra in the case of a 
        segmental reflex. e: corresponding phase spectrum. f:
        coherency spectrum.}
\end{figure}

A representative example for this model without any
reflex loop $ ( \kappa = 0 ) $, i.e.~model (\ref{nr1},\ref{nr2}),
 is given in Fig.~\ref{reflex_phase_fig2}a-c.
The non-reflex EMG peak at 10 Hz ($b_1=1.9401 , b_2=-0.9835 $) causes
an ACC peak at the same frequency. 

The ACC peak located at 5~Hz refers to the resonant behavior of
the hand ($a_1=1.9780, a_2=-0.9889 $). 
As mentioned in Section~3.3 of the companion paper (Timmer et al. 1998) the 
phase spectrum is determined by the resonance properties of the hand/muscle 
system.
\enlargethispage{2ex}
Fig.~\ref{reflex_phase_fig2}d-f displays the results 
for a delay of the reflex loop of 35 ms corresponding to a possible segmental stretch
reflex (Noth et al.~1985). $ \kappa $ was chosen equal to minus one.
\onecolumn
\twocolumn
The ACC peak caused by the resonant behavior of
the hand is shifted to a higher frequency of 6.3 Hz.
But the phase spectrum is invariant and still represents the 
characteristic features
of the mechanic properties of the hand/muscle system that have shown up
in the ACC power spectrum of Fig.~\ref {reflex_phase_fig2}a.
Therefore, a significant difference in the peak frequency
estimated from the phase spectrum and the power spectrum
indicates a contribution of the reflex to the oscillation.
There might be parameter constellations as shown in Fig.~\ref{T_von_reflex}
where a reflex does not lead to any shift of the resonance frequency.
Thus, we modify in the empirical study the mechanical properties of the system
by loading the hand with different weights to increase the probability of
detecting a possible contribution of a reflex loop.

The peak frequency is estimated from the phase spectrum by fitting
the expected theoretical phase spectrum to the empirical
one as described in Section 3.3 of the companion paper. 
The model contains a possible time delay $\Delta t$, the parameters $a_1$ 
and $a_2$ of the AR[2] process describing the properties of the hand/muscle
system. Furthermore, the twofold differentiation to obtain the ACC from the
movement $x(t)$ of the hand is considered. 
Therefore, the theoretical phase spectrum 
consists of the sum of (20,22) of the companion paper and an offset
corresponding to (21). 
The peak frequency is determined by (2,3,14) of the companion paper
from the estimated parameters $a_1$ and $a_2$.

Note that the coherency at the frequency of the reflex induced EMG 
peak increases due to the larger signal to noise ratio in accordance with (27)
of the companion paper.

\subsection {Application to measured data}

\begin{figure}[!ht]
\begin{center}
\includegraphics[scale=0.5]{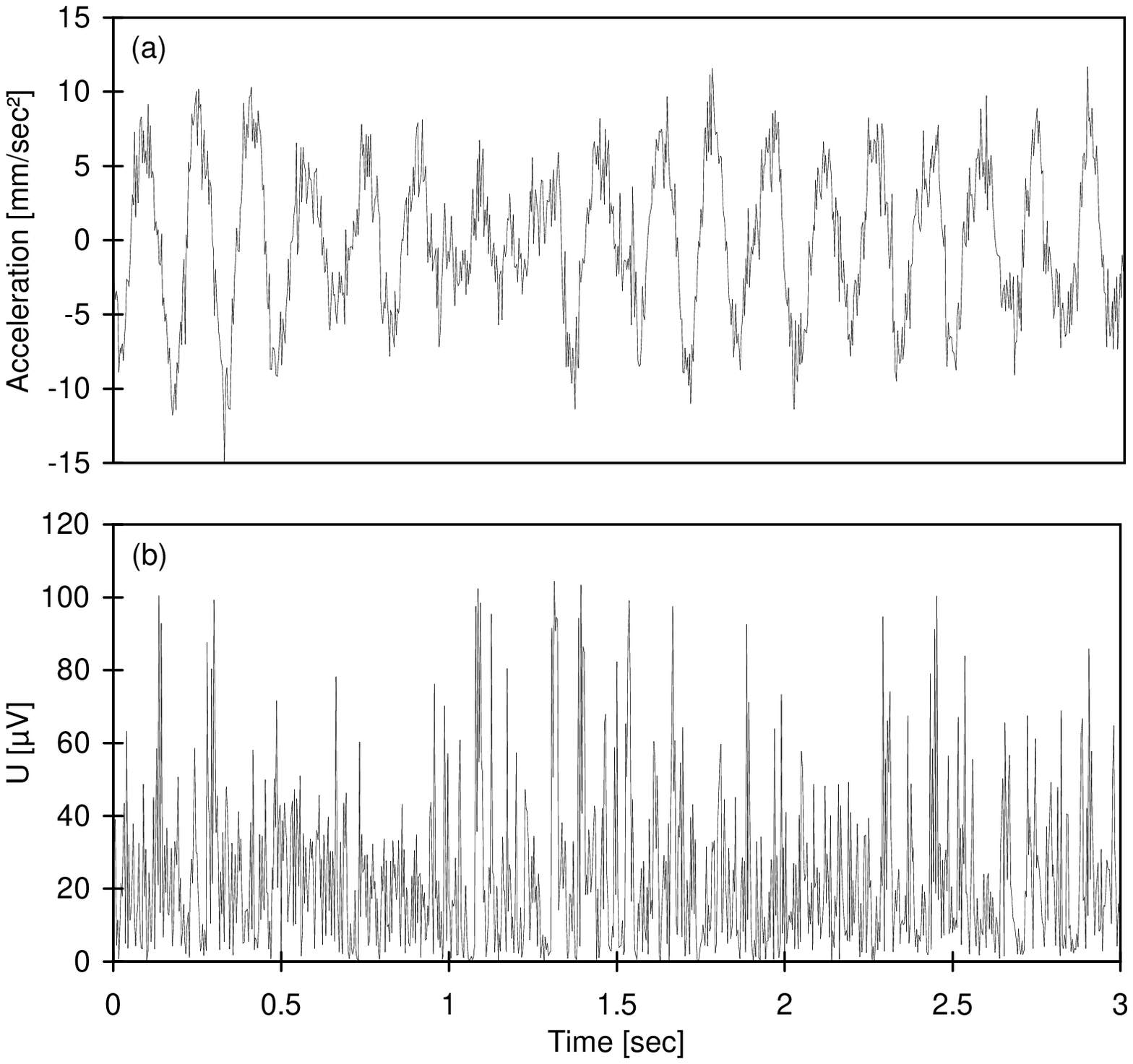}
\end{center}
\caption[]{\label{realdata1} 
Enhanced physiological hand tremor. a: Acceleration of
        the hand, b: rectified EMG}
\end{figure}

\begin{figure}[!ht]
\begin{center}
\includegraphics[scale=0.5]{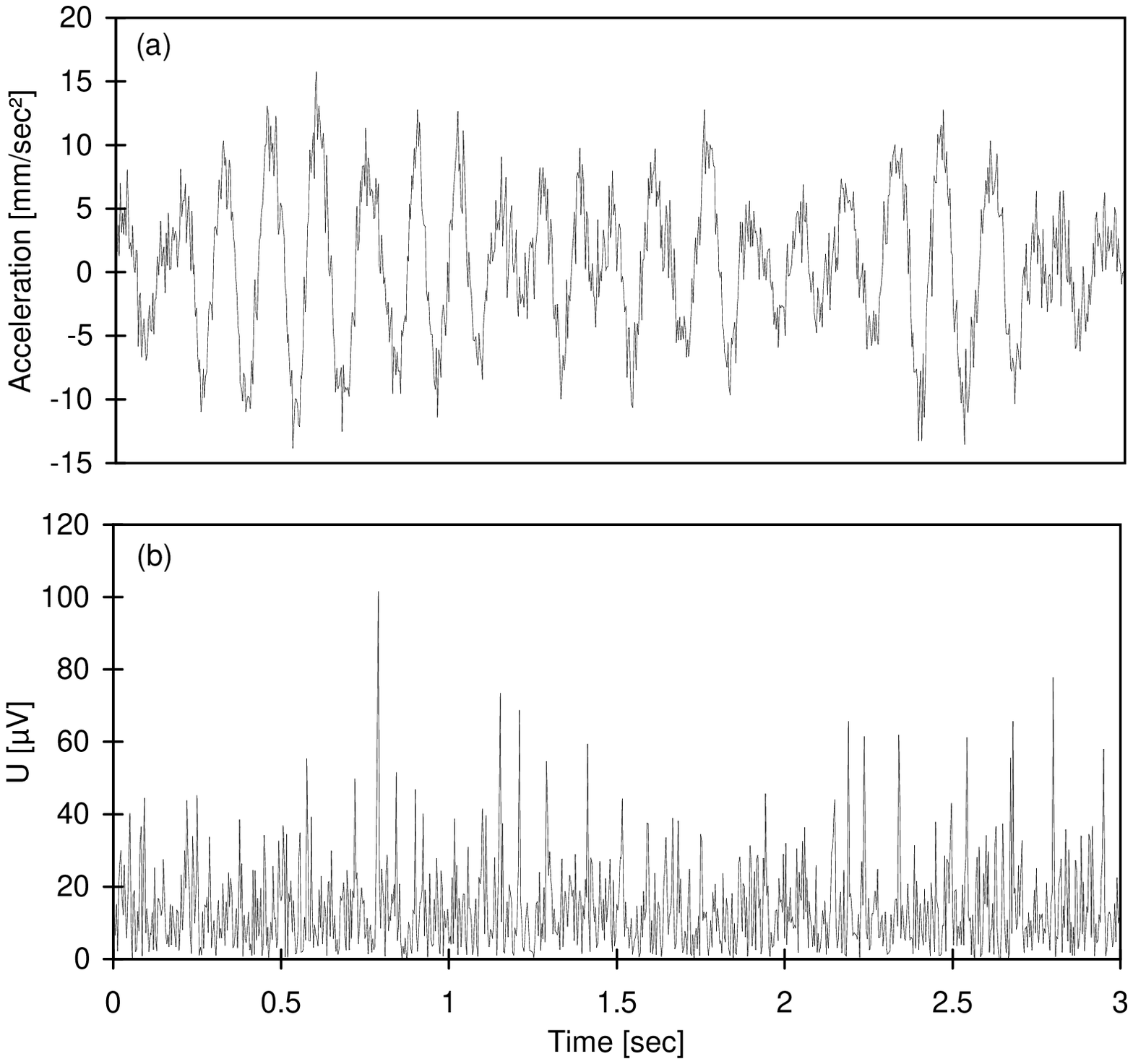}
\end{center}
\caption[]{\label{realdata2} 
Enhanced physiological hand tremor. a: Acceleration of
        the hand, b:  rectified EMG.}
\end{figure}

\begin{figure}[!ht]
\begin{center}
\includegraphics[scale=0.7]{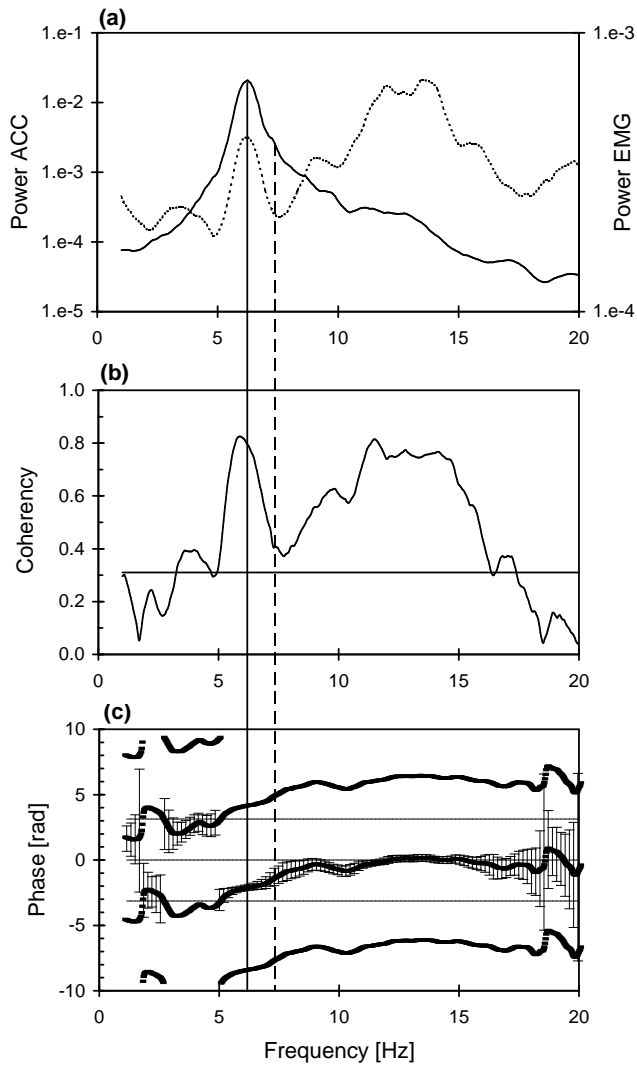}
\end{center}
\caption[]{\label{reflex_realdata1} 
 Analysis of data shown in Fig.7.
         a: power spectra (EMG: dashed line, ACC: solid line), b: coherency,
        c: phase spectrum. 
        The mechanical peak frequency estimated from the
        phase spectrum is $ 7.4 \pm 0.2 $ Hz is not consistent with that 
        estimated from the ACC power spectrum of $ 6.2 \pm 0.1$ Hz. }
\end{figure}

\begin{figure}[!ht]
\begin{center}
\includegraphics[scale=0.7]{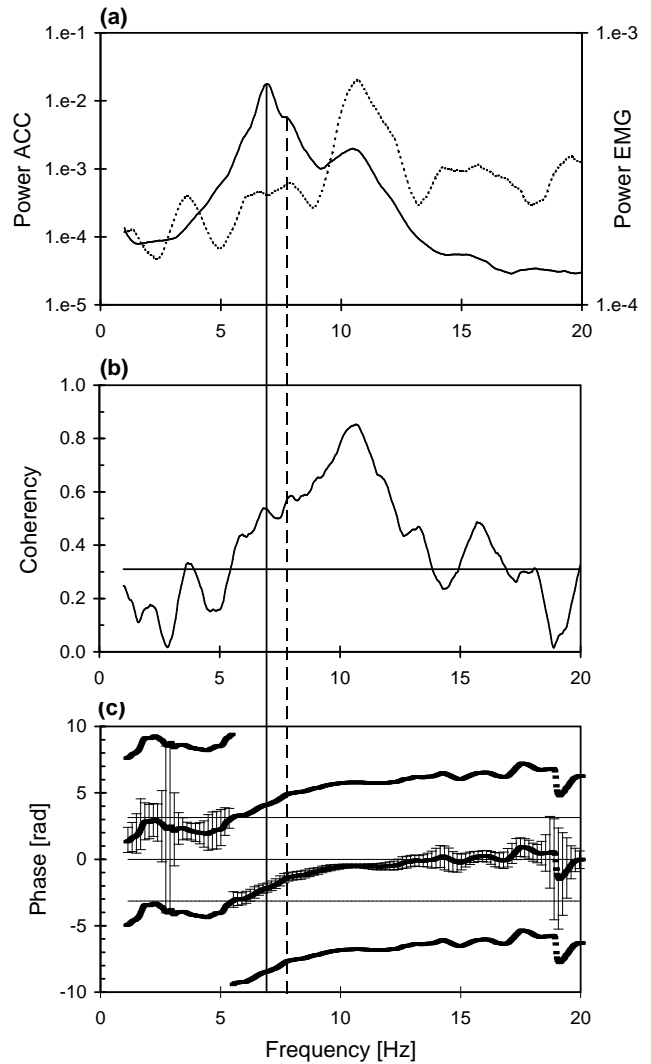}
\end{center}
\caption[]{\label{reflex_realdata2} 
Analysis of data shown in Fig.8.
         a: power spectra (EMG: dashed line, ACC: solid line), b: coherency,
        c: phase spectrum. 
        The mechanical peak frequency estimated from the
        phase spectrum is $ 7.7 \pm 0.2 $ Hz  is not consistent with 
        that estimated from the ACC power spectrum of $ 6.9 \pm 0.2 $ Hz.}
\end{figure}

Fig.~\ref{realdata1} and Fig.~\ref{realdata2} display 3 s segments
of data recorded from two subjects with ETP.
Fig.~\ref{reflex_realdata1} and \ref{reflex_realdata2} show
the results for these two examples.
In Fig.~\ref{reflex_realdata1}a the mechanical peak of the ACC
at $ 6.2 $ Hz is accompanied by a peak in the EMG, located at the same 
frequency. The neurogenic EMG peak around $ 13 $ Hz causes a shoulder in 
the ACC power spectrum according to (15) of the companion paper.
Because of the increased signal to noise ratio at the peaks, the
coherency in Fig.~\ref{reflex_realdata1}b shows significant values there.
The phase spectrum in Fig.~\ref{reflex_realdata1}c exhibits the
expected shape in dependence on the resonance properties of the hand/muscle
system. The frequency of the mechanical system estimated 
by fitting the analytical phase spectrum (20,21,22) of the companion paper
to the estimated phase spectrum is $ 7.4  \pm 0.1 $ Hz. 
The peak in the ACC power spectrum, representing the mechanical resonance
frequency, is located at $ 6.2 \pm 0.05 $ Hz (the calculation of the
confidence regions is explained in Section~3.3 of the companion paper).
Therefore, as the probability that the two frequencies are the same is less
than $0.01$, we conclude that the reflexes contribute significantly to
this tremor.

In Fig.~\ref{reflex_realdata2}a
the mechanical peak at $ 6.9 $ Hz is not accompanied by a significant
peak in the EMG. The neurogenic peak at $ 11 $ Hz causes a further
peak in the ACC power spectrum. The frequency of the mechanical
system estimated from the phase spectrum is $ 7.7 \pm 0.1 $ Hz. 
An estimation in the ACC spectrum yields to $ 6.9 \pm 0.1 $ Hz. 
Thus, again, a reflex mechanism is involved (p~$<$~0.01).
Due to the reflex contribution one would expect a peak in the EMG spectrum 
located at the mechanical ACC peak.
It might be not observable since the induced EMG activity does not 
exceed the observational noise significantly.

Because of the nonlinear structure of model
(\ref{reflex3},\ref{reflex4}) it can not be fitted directly to the
data. 

\section {Results}  \label{results}

In this section we report the results obtained by applying
methods introduced to 57 recordings of 
physiological tremor and EMG measured from 19 subjects.
For each subject the data were recorded without loading the hand,
with a load of 500~gr and 1000~gr in order to modify the resonance
frequency of the hand.

As described in Section~3.1 above and Section~3.3 of the companion paper, 
we fitted theoretical phase spectra to the empirical phase spectra
to estimate the frequency of the mechanical system. 
If no reflex mechanism was involved this frequency would 
be located at the peak frequency in the ACC power spectrum.

To obtain a good fit of the phase spectra it was always necessary to 
correct for the spurious time delay which is present because we
modeled a continuous time processes by a discrete time model due to (23) of 
the companion paper. No further time delay was found in the data. In 24 
of the 57 recordings the errors in the phase spectrum were too large to
obtain reliable results. This is caused by a poor coherency due to a small
signal-to-noise ratio. Thus, only 35 data sets were included in the analysis.

\begin{figure}
\begin{center}
\includegraphics[scale=0.45]{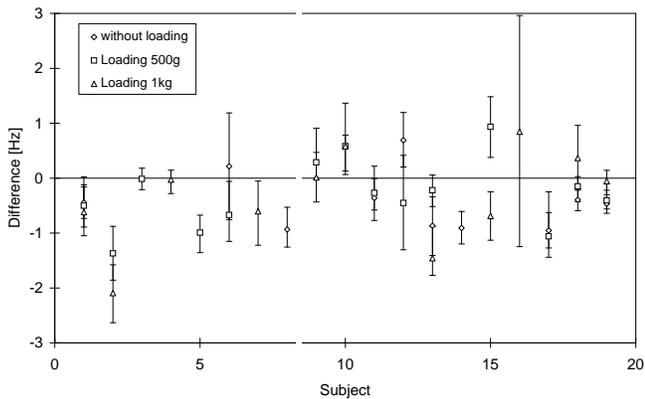}
\end{center}
\caption[]{\label{difference_fig} 
Differences in peak frequencies estimated from the phase
        spectrum and the ACC power spectrum for different loadings 
        ($\Diamond$: no loading, $\Box$: 500 gr., $\bigtriangleup$: 1000 gr.) 
        for 19 subjects with EPT.}
\end{figure}

The differences between the frequencies estimated from the phase 
and the ACC power spectrum for different loadings of the hand are given in 
Fig.~\ref{difference_fig}. 
The error-bars represent the $2\sigma$
confidence interval. Under the hypothesis that the reflex does
not contribute to the process only two of the data points 
should not be consistent with a zero difference at a significance
level of 5 \%. 19 data points are significantly different from zero.
On the one hand this result show that reflexes play a role in ETP. On the 
other hand it shows that a simple segmental stretch reflex 
is not sufficient to explain this finding. This is indicated
by the behavior of the peak frequency differences in dependence
of the loading. Simulation studies like those presented in 
Fig.~\ref {T_von_reflex} show
that the difference should become more positive under loading.
The results allow no clear cut decision on this but
the different trend suggests the involvement of more complex
reflex structures in the process, maybe a combination of different reflex
loops.

\section {Summary} \label {summary}
We investigated the relation between synchronized EMG and ACC of physiological
tremor by cross-spectral analysis. 
The phase spectrum between EMG and ACC depends only on the
mechanical properties of the hand/muscle system, i.e.~the driven part of the
system, but not on the characteristics of the driving force.
The behavior of the coherency with respect to the peaks of EMG
and ACC power spectra can be explained by its dependence on the
observational noise. Usually the maximum coherency is
located at the frequency of the EMG peak. 
Therefore, the value of the phase spectrum at the frequency of the 
EMG peak may not be interpreted as a delay. It only provides 
information about the relative location of the mechanical resonance
frequency and the synchronization frequency in the EMG. 
The time delay of reflex loops can shift the resonance peak frequency,
but do not affect the phase spectrum. 
Thus, in both cases with and without a contribution of reflexes,
the phase spectrum provides the information about the frequency 
of the mechanical peak that would show up in the absence of any reflex. 
Therefore, a significant difference in the peak frequency of the
ACC power spectrum and that calculated from the phase spectrum gives 
evidence for a contribution of reflexes to the tremor. 
In 35 time series recorded from 19 subjects with 
ETP, we found clearly that 
reflexes contribute to this tremor. The sign of the difference
was found to be opposite to that expected for segmental
reflexes from simulation studies suggesting that
more complex structures are involved.

There is no evidence in the data that reflex loops primarily cause the 
tremor. They alter the frequency, relaxation time and amplitude 
of existing oscillations to some degree. This holds for
the mechanical peak as well as for the so called neurogenic, i.e.
centrally evoked, peak. Therefore, the primary cause of ETP
is the resonant behavior of the hand and a synchronized
EMG activity that is either generated centrally or due
to the recruitment strategy of motoneurons.

\section*{References}
\begin{itemize}

\item [] Allum JHJ, Dietz V, Freund HJ (1978) Neuronal mechanisms
         underlying physiological tremor. J Neurophys 41:557-571

\item [] Allum JHJ (1984) Segmental reflex, muscle mechanical and
         central mechanisms underlying human physiological tremor.
         In: Findley LJ Capildeo R (Eds.) Movement disorders: Tremor. 
         London. Macmillan, pp. 135-155 

\item [] Bawa P, Mannard A, Stein RB (1976a) Effects of elastic loads on the 
         contraction of cat muscles. Biol Cybern 22:129-137

\item [] Bawa P, Mannard A, Stein RB (1976b) Predictions and experimental
         tests of a visco-elastic muslce model using elastic and
         inertal loads. Biol Cybern 22:139-145

\item [] Christakos CN (1982) A study of the electromyogram using a population
         stochastic model of skeletal muscle. Biol Cybern 45:5-12

\item [] Elble RJ, Koller WC (1990) Tremor. The Johns Hopkins
         University Press, Baltimore
 
\item [] Elble RJ, Randall JE (1976) Motor-unit activity responsible
         for 8- to 12-Hz component of human physiological finger tremor. 
         J Neurophys 39:370-383

\item [] Fox JR, Randall JE (1970) Relationship between forearm tremor
         and the biceps electromyogram. J Appl Neurophys 29:103-108

\item [] Gantert C, Honerkamp J, Timmer J (1992) Analyzing the
         dynamics of hand tremor time series. Biol Cybern 66:479-484

\item [] Hagbarth KE, Young RR (1979) Participation of the stretch reflex
         in human physiological tremor. Brain 102:509-526

\item [] H\"omberg V, Hefter H, Reiners K, Freund HJ (1987) Differential
         effects of changes in mechanical limb properties on physiological
         and pathological tremor. J Neuro Neurosurg Psychiatry 50:568-579

\item [] Iaizzo PA, Pozos RS (1992) Analysis of multiple EMG and 
         acceleration signals of various record length as a means to study
         pathological and physiological oscillations. Electromyogr clin
         Neurophysiol 32:359-367

\item [] Lippold OCJ (1957) The rhythmical activity of groups of motor
         units in the voluntary contraction of muscle. J Physiol 137:473-487

\item [] Lippold OCJ (1970) Oscillation in the stretch reflex arc and the
         origin of the rhythmical 8-12 c/s component of physiological
         tremor. J Physiol 206:359-382

\item [] Lippold OCJ (1971) Physiological tremor. Scientific American. 65-73

\item [] Marmarelis VZ (1989) Volterra--Wiener Analysis of a class of
         nonlinear feedback systems and application to sensory
         biosystems . In:
         Physiological System Modeling, Ed.: VZ Marmarelis. New York,
         Plenum Press. pp. 1-52 

\item [] Marsden CD (1984) Origins of normal and pathological tremor. In:
         Findley LJ Capildeo R (Eds.) Movement disorders: Tremor. 
         London. Macmillan, pp. 37-84

\item [] Matthew PBC (1994) The simple frequency response of human stretch
         reflexes in which either short- or long-latency components
         predominate. J Physiol 481:777-789

\item [] Noth J, Podoll K, Friedemann H (1985) Long-loop reflexes in
         small hand muscles studied in normal subjects and in patients
         with Huntington's disease. Brain 108:65-80

\item [] Oguzt\"oreli MN, Stein RB (1975) An analysis of oscillations 
         in neuro-muscular systems. J Math Biol 2:87-105

\item [] Oguzt\"oreli MN, Stein RB (1976) The effects of multiple 
         reflex pathways on the oscillations in neuro-muscular systems.
         J Math Biol 3:87-101

\item [] Oguzt\"oreli MN, Stein RB (1979) Interactions between centrally 
         and peripherally generated neuromuscular oscillations.
         J Math Biol 7:1-30

\item [] Pashda SM, Stein RB (1973) The bases of tremor during a
         maintained posture. In: Control of Posture and Locomotion, Ed.: RB
         Stein, KG Pearson, RS Smith, JB Redford. New York. Plenum.
         Press  pp 415-419

\item [] Randall JE (1973) A Stochastic Time Series Model for Hand
         Tremor. J Appl Physiol 34:390-395

\item [] Rack PMH, Ross HF, Brown TIH (1978) Reflex response during 
         sinusoidal movement of human limbs. In: Cerebral motor control
         in man: Long loop mechanisms. Prog. clin. Neurophys. Vol 4, 
         Ed.: JE Desmedt. Karger, Basel. pp. 216-228

\item [] Rietz RR, Stiles RN (1974) A viscoelastic mass mechanism as a basis 
         for normal postural tremor. J Appl Physiol 37:852-860 

\item [] Stein RB, Oguzt\"oreli MN (1975) Tremor and other oscillations 
         in neuromuscular systems. Biol Cybern 22:147-157

\item [] Stein RB, Oguzt\"oreli MN (1978) Reflex involvement in the 
         generation and control of tremor and clonus. In: Physiological tremor,
         pathological tremor and clonus. Prog. clin. Neurophys. Vol 5, 
         Ed.: JE Desmedt. Karger, Basel. pp. 28-50

\item [] Stiles NS, Randall JE (1967) Mechanical factors in human tremor
         frequency. J Appl Physiol 23:324-330

\item [] Stiles NS (1980) Mechanical and Neural feedback factors in
         postural hand tremor of normal subjects. J Neurophys 44:40-59

\item [] Timmer J, Lauk M, Pfleger W, Deuschl G (1998)
         Cross-spectral analysis of physiological tremor and muscle 
         activity. I. Theory and application to unsychronized EMG.
         submitted to Biol Cybern. 

\item [] Young RR, Hagbarth KE (1980) Physiological tremor enhanced
         by manoeuvres affecting the segmental stretch reflex. J Phys
         43:248-256

\end {itemize}
\end {document}